\documentclass[journal]{IEEEtran}

\usepackage{graphicx} 
\usepackage{epstopdf} 
\usepackage{subfigure} 
\usepackage{cite}
\usepackage{multirow} 
\usepackage{enumerate}
\usepackage{amsmath} 
\usepackage{amssymb} 
\usepackage{amsthm} 
\usepackage{algorithm} 
\usepackage{algorithmic}

\renewcommand{\algorithmicrequire}{\textbf{Input:}}
\renewcommand{\algorithmicensure}{\textbf{Output:}}

\begin{document}

\title{Parallel Scheduling Algorithm based on Complex Coloring for Input-Queued Switches}

\author{Lingkang~Wang,
        Tong~Ye,~\IEEEmembership{Member,~IEEE,}
        Tony~T.~Lee,~\IEEEmembership{Fellow,~IEEE,}
        and~Weisheng~Hu,~\IEEEmembership{Member,~IEEE}
\thanks{This work was supported by the National Science Foundation of China (61271215, 61172065, and 61433009).}
\thanks{The authors are with the State Key Laboratory of Advanced Optical Communication Systems and Networks, Shanghai Jiao Tong University, Shanghai 200240, China (e-mail: \{sin7cera, yetong, ttlee, wshu\}@sjtu.edu.cn).}}

\maketitle

\begin{abstract}
This paper explores the application of a new algebraic method of edge coloring, called complex coloring, to the scheduling problems of input queued switches. The proposed distributed parallel scheduling algorithm possesses two important features: optimality and rearrangeability. Optimality ensures that the algorithm always returns a proper coloring with the minimum number of required colors, and rearrangeability allows partially re-coloring the existing connection patterns if the underlying graph only changes slightly. The running time of the proposed scheduling algorithm is on the order of $O(\log^2 N)$ per frame, and the amortized time complexity, the time to compute a matching per timeslot, is only $O(\log N)$. The scheduling algorithm is highly robust in the face of traffic fluctuations. Since the higher the variable density, the higher the efficiency of the variable elimination process, complex coloring provides a natural adaptive solution to non-uniform input traffic patterns. The proposed scheduling algorithm for packet switching can achieve nearly 100\% throughput.
\end{abstract}


%
\IEEEpeerreviewmaketitle

\section{Introduction}\label{intro}
%
%
%
%
\IEEEPARstart{T}{he} core issue of all input-queued (IQ) switches is the buffering and scheduling of input traffic. Most switching systems in practical applications are networks with input-queues \cite{MarsanTON02,whitepaper09,whitepaper10}. It is well-known that the throughput of the IQ switch is limited by output contentions. To maximize the throughput, a traffic scheduler is indispensable to properly set up a connection pattern for the switch fabric in each timeslot to avoid output contentions. The best existing scheduling algorithm \cite{iSLIP} is still a far cry from the capacity of today’s datacenters, in which hundreds of thousands of servers are interconnected by large numbers of switches. Driven by the recent surge of traffic generated by datacenters, it is expected that a switch would provide more than 1 Pb/s throughput. In other words, the number of ports of the switch would increase to more than 1000 and the line rate of each port could be higher than 1 Tb/s \cite{refOIDA}. 

A connection pattern of an $N{\times}N$ switch in a timeslot is a matching of an $N{\times}N$ bipartite graph \cite{survey03}. For packet switching, the scheduling algorithm computes a matching in every timeslot, or a set of matchings in every several timeslots, called a frame, which corresponds to the edge coloring of an $N{\times}N$ bipartite graph \cite{Focs03}. The minimum number of colors required to color the edges of a graph $G=(V,E)$, with vertex set $V$ and edge set $E$, such that no adjacent edges have the same color is called the chromatic index $\chi_e(G)$ of graph $G$. A theorem proved by Vizing \cite{Vizing's} states that the chromatic index is either $\Delta$ or $\Delta+1$, where $\Delta$ is the maximum vertex degree of graph $G$. As a corollary of the Hall’s theorem \cite{Hall's}, the chromatic index of a bipartite graph $G$ is $\chi_e(G)=\Delta$, thus bipartite graphs are all $\Delta$-edge-colorable. 

In graph theory, the proof of Vizing’s theorem and the construction of the Edmonds’ blossom-shrinking algorithm for finding a maximum matching \cite{graphIntro02} are all based on two-colored subgraphs, called alternating paths. The original color-exchange method was devised by Alfred Kempe in his endeavor to prove the four-color theorem \cite{Kempe1879}. The basic idea of Kempe’s method is to resolve color conflicts by exchanging vertex colors of two-colored subgraphs.

The complex coloring proposed in \cite{complexColor} extends the concept of Kempe chain to color-exchange operations performed on two-colored alternating paths. In a nutshell, the complex coloring is an algebraic method of variable eliminations to color edges. Each edge $e{\in}E$ is considered as a pair of links; each link is a half-edge. Let $C$ be the set of colors. The complex coloring of a graph $G=(V,E)$ assigns a color pair, or a complex color, to each edge $e{\in}E$, one color assigned for each link. Suppose that the complex color assigned to an edge $e$ is $c(e)=(\alpha,\beta),\alpha,\beta{\in}C$, then the colored edge $e$ is a variable if $\alpha{\neq}\beta$; otherwise, $c(e)=(\alpha,\alpha)$ is a constant for any color $\alpha$. A proper $\Delta$-edge-coloring of graph $G$ is achieved by eliminating all variables. 

The aim of this paper is to develop an ultra-fast distributed parallel scheduling algorithm based on complex coloring. The parallelized complex coloring possesses two important features: optimality and rearrangeability. Optimality ensures that the algorithm always returns a proper coloring with the minimum number of required colors, and rearrangeability allows partially re-coloring the existing connection patterns if the underlying graph only changes slightly.


 

\subsection{Previous Works}\label{relatedWork}
To achieve better system utilization and adapt to input traffic fluctuation, on-line scheduling algorithms are commonly used to compute connection patterns on a slot-by-slot basis in real time. A class of scheduling algorithms was proposed based on the maximum size matching (MSM) algorithm \cite{iSLIP,karpMSM,PIM,DRRM1998,DRRM01,FIRM,Mneimneh08,SRRM}. The goal of the algorithms is to maximize the number of connections from inputs to outputs in each timeslot, and thus maximize instantaneous bandwidth utilization \cite{iSLIP}. Currently, the most efficient algorithm takes $O(N^{2.5})$ time \cite{karpMSM}. It is too complex to be implemented in practice and it may cause some input queues to be starved of service indefinitely \cite{iSLIP}, which will make the system unstable. To reduce the computational complexity, several heuristic iterative algorithms \cite{iSLIP,PIM,DRRM1998,DRRM01,FIRM,Mneimneh08,SRRM}, such as PIM \cite{PIM}, \emph{i}SLIP \cite{iSLIP},  and DRRM \cite{DRRM1998,DRRM01}, were proposed to approximate the MSM. Most of these algorithms have a time complexity of $O(N\log N)$, which is not fast enough for real time on-line implementation in current large-scale and high-speed switches \cite{Keslassy02}. Moreover, the MSM-based algorithms only consider whether each VOQ is empty during the computation of matching; therefore, those algorithms cannot sustain a stable throughput under non-uniform traffic \cite{withoutSpeedup,MSMunstable}, which is the typical scenario in current applications of datacenter networks.

Based on the idea of maximum weighted matching (MWM), scheduling algorithms were designed in \cite{iLQF,Mckeown92,RPA,APSARA03,DLQF} to guarantee a stable throughput under non-uniform traffic. Different from the MSM-based scheduling algorithms, the MWM-based scheduling algorithms take the system statistic information, such as the queue length or the delay of the head-of-line packet of each VOQ, into consideration in the calculation of a matching. Thus, the MWM-based algorithm can achieve 100\% throughput even under non-uniform traffic. However, the MWM-based algorithms are typically too complex to be implemented in real time. As the study reported in \cite{RPA} shows, the computational complexity of two distributed algorithms, \emph{i}LQF and \emph{i}OCF, is on the order of $O(N^2\log⁡N)$. Although the most efficient algorithm reported in \cite{APSARA03} can converge to a solution on the order of $O(N)$ running time, it is a centralized algorithm and requires the collection of statistic information of all input ports in each timeslot.

Frame-based scheduling \cite{frameMatching,Capturedframe,Chao11,BinWu09,logDelay,frameGame,Focs03} is an alternative way to cope with non-uniform traffic. In these schemes, a batch of consecutive timeslots forms a time frame, and the packets that arrive in a frame are collectively scheduled by the scheduler at the same time. Because a time frame provides a snapshot of the system statistic, the frame-based scheduling is quite robust in dealing with non-uniform traffic, even using simple coloring algorithms. To use this advantage, the iSLIP and the PIM were generalized to frame-based scheduling algorithms in \cite{frameMatching,Capturedframe}, respectively, which have a time complexity on the order of $O(N\log⁡N)$ per timeslot, and can achieve a high throughput under non-uniform traffic. On the other hand, a randomized frame-based scheduling algorithm was proposed in \cite{Focs03}, which colors the bipartite graph using a random coloring method. The complexity per timeslot of such an algorithm is at least $O(N)$, which is not fast enough for current applications. For example, at a 100-Gbps line rate, a 64-byte packet may last around 5ns, while the computation time for a switch with 1000 input ports would be on the order of 10ns. 

To avoid on-line computation while providing the bandwidth guarantee for each input/output pair, a quasi-static scheduling algorithm, called path switching, was proposed in \cite{pathSwitching}, which was later called Birkhoff-von Neumann (BvN) switching in \cite{BvN}. This scheme achieves the bandwidth guarantee by repeatedly running a set of predetermined connection permutations, which are calculated from the average loading of all input/output pairs. The on-line complexity of this algorithm is $O(1)$ since it does not need to compute the permutations on-the-fly. Though this algorithm is attractive in terms of on-line complexity, it requires timely updating traffic statistics, which could be quite difficult in practice especially when the input traffic is highly bursty. 

To deal with unpredictable traffic fluctuations, a two-stage load-balanced BvN (LB-BvN) switch is proposed in \cite{LB_BvN}, where a set of $N$ circular-shift permutations is periodically running in both stages. In the first stage, the input traffic is evenly distributed over all input ports of the second stage, while the second stage delivers the packets to their destinations. Despite that the LB-BvN switch can easily adapt to traffic fluctuation \cite{Keslassy02,ZZhang11}, immoderate traffic dispatching in the first stage inevitably induces a severe packet out-of-sequence problem at the outputs. A lot of previous work \cite{JJJaramillo06,XWang08,BHu10,ZZhang11} has shown that the out-of-sequence problem can be solved, however, at the expense of a high computation or hardware complexity.

All in all, it remains a challenge to find efficient scheduling algorithms with high performance and low complexity for current large-scale high speed switches.

\subsection{Summary of our work}\label{ourWork}
The fundamental principle of complex coloring is variable elimination, and the basic operation is the color exchange between two adjacent edges. However, performing simultaneous color-exchange operations on the two end vertices of an edge is prohibited because of the possibility of creating new variables. In a bipartite graph $G=(X{\cup}Y,E)$, the vertices in set $X$ are non-adjacent and so are the vertices in $Y$. If simultaneous color-exchange operations are performed on the vertices in set $X$ and set $Y$ alternately, then parallel processing of complex coloring in a distributed manner becomes viable. We aim at developing an ultra-fast scheduling algorithm based on this parallelization property of bipartite graphs. Moreover, a bipartite graph does not have odd cycles, which ensures that all variables can be eliminated, and a proper edge coloring with a minimal number of colors can be obtained.

The parallel complex coloring, however, may introduce deadlock situations in the process, when variable edges are trapped in infinite loops. We establish a stopping rule for parallel processing to prevent the process running indefinitely. Fortunately, the number of edges involved in deadlocks is very small. Therefore, we propose an on-line scheduling algorithm for frame-based packet switching, where the remaining variable edges will simply be ignored in current time frame and stay down with the bipartite graph of the next time frame. Since the number of remaining uncolored edges is very small in each time frame, this strategy only sacrifices a tiny fraction of bandwidth, and achieves nearly 100\% throughput.

In summary, our contribution in this paper is to synthesize an ultrafast scheduling algorithm based on distributed parallel processing of complex coloring that achieves the following objectives:
\begin{itemize}
\item
\emph{Low Complexity}: The running time of our algorithm for each frame is $O(\log^2 N)$, and the amortized time complexity per timeslot is only $O(\log N)$. Throughout this paper, the amortized time complexity is referred to the required time to compute a matching in a timeslot.
\item
\emph{High Throughput}: Our algorithm can achieve nearly 100\% throughput.
\item
\emph{Robustness}: The higher the variable density, the higher the efficiency of the variable elimination process. Therefore, complex coloring provides a natural adaptive solution to schedule non-uniform input traffic patterns.
\end{itemize}

The rest of the paper is organized as follows. In Section \ref{preliminary}, we introduce the scheduling of frame-based packet switching and then describe the formulation of scheduling problems as edge coloring of bipartite graphs. Next, we present a brief description of the variable elimination process of complex coloring. Section \ref{parallelCColoring} discusses issues related to parallel processing of complex coloring, focusing on deadlocks and stopping rules. In Section \ref{schedulingAlgorithms}, we propose an on-line scheduling algorithm for frame-based packet switching. Section \ref{performance} discusses performance evaluations of the proposed scheduling algorithm under different traffic patterns. Section \ref{conclusion} draws the conclusion of this paper.

\section{Preliminaries of Scheduling and Coloring}\label{preliminary}
The fundamental principle of numerous switching scheduling algorithms is based on edge coloring of bipartite graphs. In this section, we introduce the basic concepts and terminologies of frame-based on-line scheduling for packet switches. The purpose is to establish the correspondence between the physical aspects of this scheduled switching system and its respective counterpart in the edge coloring formulation of scheduling problems. Next, we briefly describe a newly proposed algebraic method \cite{complexColor}, called complex coloring, for edge coloring of bipartite graphs. From the well-known property that a bipartite graph only contains even cycles, we show that the complex coloring method can always produce a proper coloring of the bipartite graph with the minimum number of colors that equals the maximum degree of the graph.

\subsection{On-line Scheduling}\label{onlineIntro}
For an internally non-blocking packet switch with input buffers, on-line scheduling algorithms calculate the connection pattern of the switch fabric in real time for arrival packets to avoid output contention. In this paper, we consider the general frame-based packet scheduling for an $N{\times}N$ input-queued non-blocking switch, such as a crossbar switch. We assume that time is slotted and the packet size is fixed, and at most one packet can be sent from each input and at most one packet can be received by each output in one timeslot. A frame-based scheduling algorithm constructs proper connection patterns for input packets in three stages: accumulating stage, scheduling stage, and switching stage. A batch of $f$ consecutive timeslots forms a time frame and each stage lasts one frame. In the first stage, the arrival packets are accumulated in the input buffer to form a frame. In the second stage, the scheduler calculates a set of $f$ permutations for packets accumulated in a frame. In the third stage, the packets are switched to their destinations according to the scheduled permutations. The three stages of the frame-based scheduling process can be implemented in a pipelining manner \cite{BinWu09}. At the time when the scheduler calculates the permutations for current frame, the switch sends the previous frame of packets to the outputs and accumulates new arrivals to form the next frame of packets.

In an $N{\times}N$ input queued switch, contentions among packets in each time frame can be represented by a bipartite graph $G=(X{\cup}Y,E)$, where the vertex $x_i{\in}X$ and the vertex $y_j{\in}Y$ denote input port $i$ and output port $j$, respectively, for $i,j=1,2,\dots,N$. An edge $e(x_i,y_j){\in}E$ between the two vertices $x_i$ and $y_j$ represents a packet to be switched from input port $i$ to output port $j$. The degree of input vertex $x_i$ is the number of packets to be sent from input port $i$ in a frame, and the degree of output vertex $y_j$ is the number of packets destined for output port $j$. Hence, the maximum degree $\Delta$ of bipartite graph $G$ equals frame size $f$ if the number of packets destined to each output port is no more than the frame size.

Let $C$ be the set of colors. Suppose that an edge colored by the color $c_t{\in}C$ represents a connection in timeslot $t$ of the frame, then the frame-based scheduling is identical to the edge coloring of graph $G$. That is, the edges colored by the same color constitute a matching, which corresponds to a connection pattern of the input-queued switch. Since at most one packet can be sent from each input and at most one packet can be received by each output in one timeslot, the two edges incident to the same vertex in graph $G$ must be colored with distinct colors. 

\begin{figure}[!t]
\centering
\includegraphics[scale=0.75]{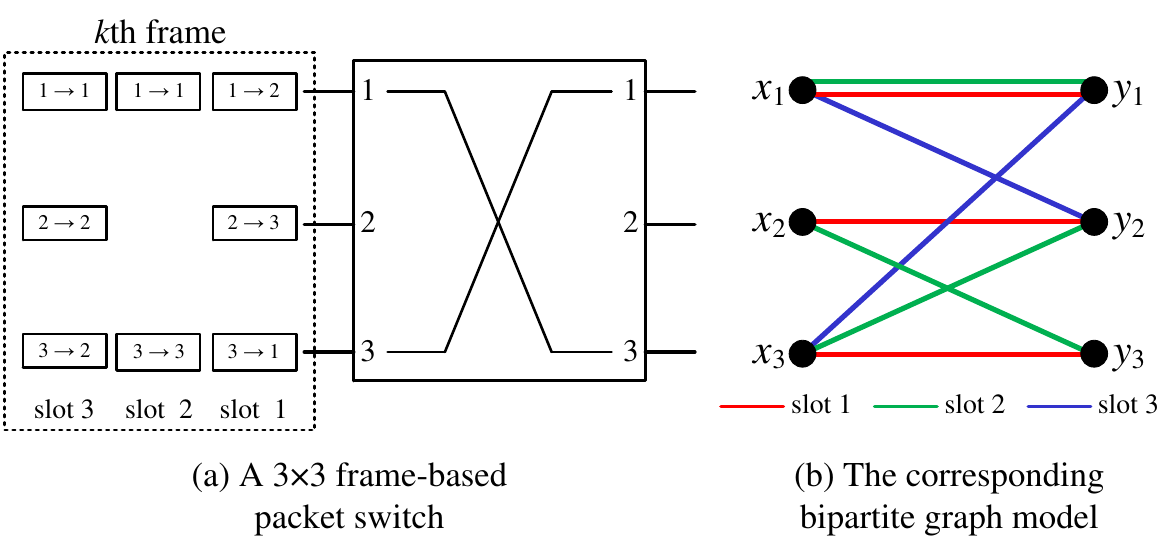}
\caption{Correspondence between the frame-based scheduling and the edge coloring.}
\label{switch_bipartite}
\end{figure}

An example of frame-based scheduling of a $3{\times}3$ packet switch with frame size $f=3$ is shown in Fig.~\ref{switch_bipartite}. Note that multiple edges between two adjacent vertices in graph $G$ are allowed because the number of packets from an input $i$ to an output $j$ in each frame may be more than one. An immediate consequence of the coloring formulation is that the number of colors in $C$, which equals frame size $f$, should not be less than the maximum degree $\Delta$ of graph $G$. In edge coloring of bipartite graph, it is well-known that the minimum required number of colors equals the maximum degree of the graph \cite{graphIntro02}. Therefore, a scheduling is optimal if the number of colors in $C$ equals the maximum degree $\Delta$ of graph $G$. 

\subsection{Complex Coloring of Bipartite Graph}\label{CCIntro}
\begin{figure}[h]
\centering
\includegraphics[scale=0.9]{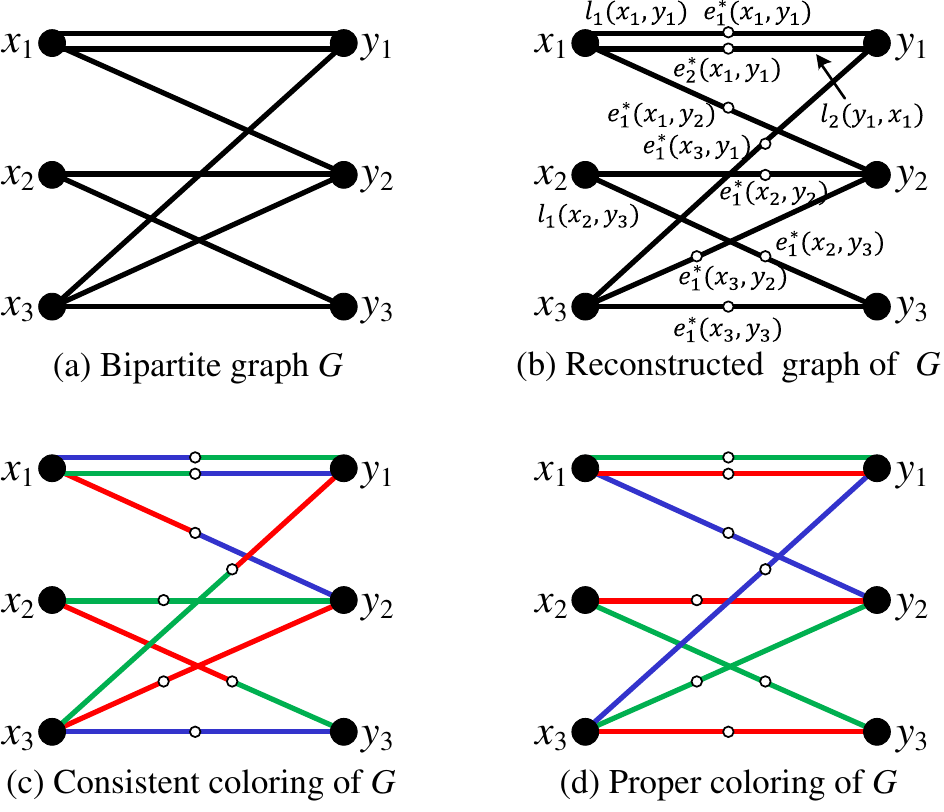}
\caption{Complex coloring of a bipartite graph $G$.}
\label{CCconcept}
\end{figure}

The complex coloring is an algebraic method proposed in \cite{complexColor}, which solves the edge-coloring problem by using a variable elimination process. In a bipartite graph $G=(X{\cup}Y,E)$, suppose that a fictitious vertex is inserted in the middle of an edge $e_r(x_i,y_j){\in}E$, in which the fictitious vertex is denoted as $e_r^*(x_i,y_j)$, then the edge $e_r(x_i,y_j)$ is divided into two links, expressed by $l_r(x_i,y_j)$ and $l_r(y_j,x_i)$. These two links connect the fictitious vertex $e_r^*(x_i,y_j)$ to the two end vertices $x_i$ or $y_j$, respectively. As an example, the reconstructed graph of the bipartite graph displayed in Fig.~\ref{CCconcept}(a) is shown in (b).

Instead of coloring the edges, the complex coloring method starts assigning colors to links. The two links of an edge could be colored by using different colors, but the colors of those links incident to the same node must be different. We adopt the following notations defined in \cite{complexColor}. 
\newtheorem{def1}{Definition}
\begin{def1}
The color of an edge $e_r(x_i,y_j)$ is a two-tuple color vector, denoted as $\vec{e}_r(x_i,y_j)=(c_r(x_i,y_j),c_r(y_j,x_i))$, which represents the color $c_r(x_i,y_j)$ of link $l_r(x_i,y_j)$ and $c_r(y_j,x_i)$ of link $l_r(u_j,v_i)$. The colored edge $\vec{e}_r(x_i,y_j)$ is a variable if $c_r(x_i,y_j){\neq}c_r(y_j,x_i)$. Otherwise, $\vec{e}_r(x_i,y_j)$ is a constant colored edge.
\end{def1}

The coloring is consistent if the colors assigned to the links incident to the same vertex are all distinct. For a bipartite graph $G$ with maximum degree $\Delta$, it is easy to show that graph $G$ can be consistently colored by a set of $\Delta$ colors. For example, the reconstructed graph with maximum degree $\Delta=3$ shown in Fig.~\ref{CCconcept}(b) is consistently colored by the set of colors $C=\{r,g,b\}$, as depicted in Fig.~\ref{CCconcept}(c). The graph is properly colored if the coloring is consistent and all edges are constant colored edges. A proper coloring of graph $G$ is illustrated in Fig.~\ref{CCconcept}(d).

Suppose graph $G$ is consistently colored by complex colors, then a proper coloring of graph $G$ can be obtained by variable eliminations, provided that the consistency of coloring is preserved throughout the variables elimination process. The consistent variable elimination is based on the binary color-exchange operation performed on two adjacent colored edges defined as follows.
\newtheorem{def2}[def1]{Definition}
\begin{def2}
Let $(c_r(x_i,y_j),c_r(y_j,x_i)){\circ}(c_{r'}(y_j,x_k),c_{r'}(x_k,y_j))$ denote the complex colors of two adjacent colored edges $\vec{e}_r(x_i,y_j)$ and $\vec{e}_{r'}(y_j,x_k)$, where $c_r(x_i,y_j)=b$, $c_{r'}(y_j,x_k)=a$ and $a,b{\in}C,a{\neq}b$. The binary operation $\otimes$ of color exchange is defined as follows:
\begin{align}
(c_r(x_i,y_j),c_r(y_j,x_i)){\otimes}(c_{r'}(y_j,x_k),c_{r'}(x_k,y_j)) \nonumber\\
=(c_r(x_i,y_j),b){\otimes}(a,c_{r'}(x_k,y_j)) \nonumber\\
{\Rightarrow}(c_r(x_i,y_j),a){\circ}(b,c_{r'}(x_k,y_j)). \nonumber
\end{align}
\end{def2}

\begin{figure*}[!t]
\centering
\includegraphics[scale=0.81]{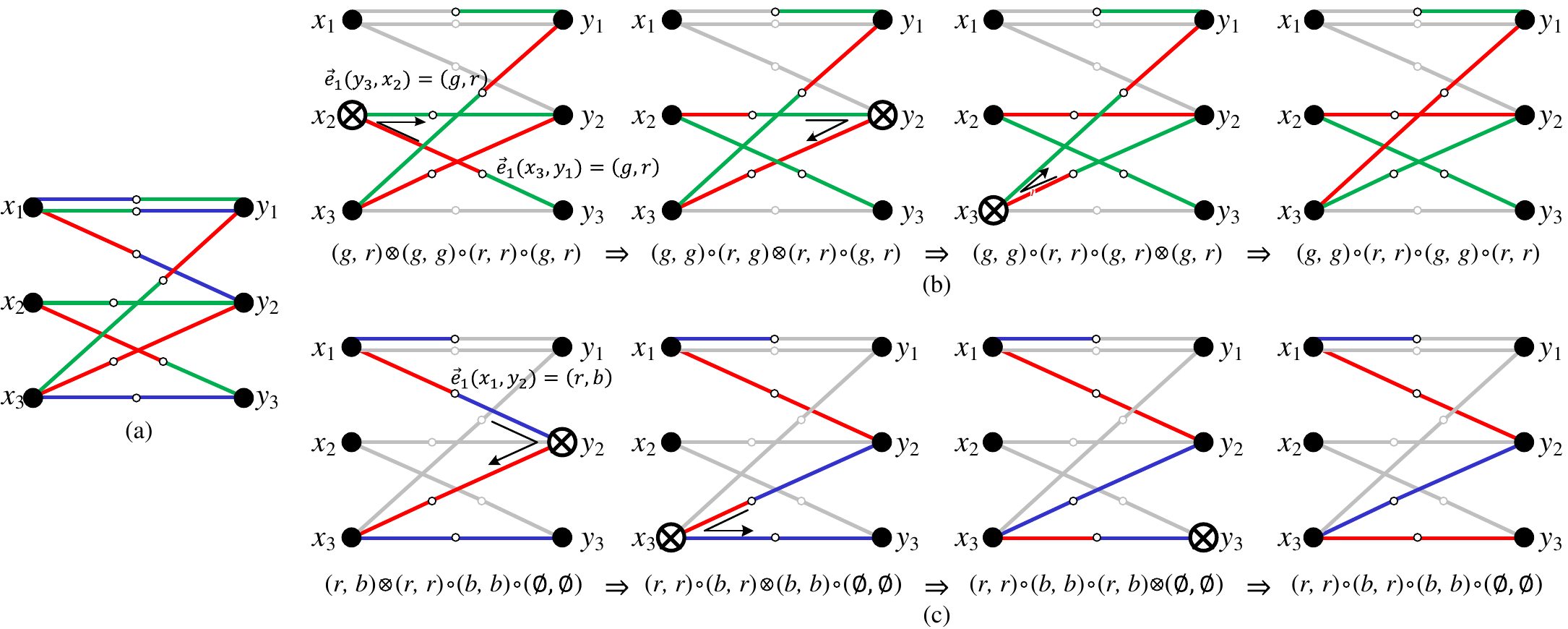}
\caption{Variable elimination via Kempe walks.}
\label{KempeWalk}
\end{figure*}

The color-exchange operation can either eliminate adjacent variables or introduce new variables. A color-exchange operation is \emph{effective} if it does not increase the number of variables. For example, in Fig.~\ref{KempeWalk}(c), the color-exchange operation $(c_1(y_1,x_1),c_1(x_1,y_1)){\otimes}(c_1 (x_1,y_2 ),c_1 (y_2,x_1 ))=(g,b){\otimes}(r,b)=(g,r){\otimes}(b,b)$  will eliminate both variables and produce a variable $(g,r)$ and a constant $(b,b)$, thus it is an effective exchange. In the complex coloring, only the effective color-exchange operations are performed.

Variables that are separated in a consistently colored bipartite graph $G$ must be moved next to each other before they can be eliminated. To preserve the consistency of complex coloring, a variable with complex color $(a,b)$ is only allowed to move within a two-colored subgraph, called $(a,b)$-Kempe path, that is colored by color $a$ and $b$. The sequence of effective color exchanges that move the variable is called a Kempe walk. The definition of a Kempe path is formally given as follows.
\newtheorem{def3}[def1]{Definition}
\begin{def3}
A $(a,b)$-Kempe path, or simply $(a,b)$ path, where $a,b{\in}C$ and $a{\neq}b$, is a sequence of adjacent links $l_1,l_2,{\dots},l_n$, where $c(l_i){\in}\{a,b\}$ for $i=1,2,{\dots},n$. The vertices on the path are called the interior chain of the path. There are two types of maximum $(a,b)$ path:
\begin{enumerate}[{1)}]
\item
$(a,b)$ cycle: the two end-links are adjacent to each other;
\item
$(a,b)$ open path: the two end-links are adjacent to only one $a$ or $b$ colored link.
\end{enumerate}
\end{def3}

As an example, in Fig.~\ref{KempeWalk}(a), the two-colored subgraph $l_1(x_1,y_1)$, $l_1(y_1,x_1)$, $l_2(y_1,x_1)$, $l_2(x_1,y_1)$ is a $(b,g)$ cycle, and the subgraph $l_1(x_1,y_1)$, $l_1(x_1,y_2)$, $l_1(y_2,x_1)$, $l_1(y_2,x_3)$, $l_1(x_3,y_2)$, $l_1(x_3,y_3)$, $l_1(y_3,x_3)$ is a $(b,r)$ open path.

The Kempe walk of moving an $(a,b)$ variable on an $(a,b)$ path is a sequence of effective color-exchange operations performed on its interior chain. Fig.~\ref{KempeWalk}(b) illustrates the Kempe walk that moves the colored variable $\vec{e}_1(y_3,x_2)=(g,r)$ to the variable $\vec{e}_1(x_3,y_1)=(g,r)$ by performing the following sequence of color exchanges on its interior chain:
\begin{align}
(g,r)&{\otimes}(g,g){\circ}(r,r){\circ}(g,r){\Rightarrow}(g,g){\circ}(r,g){\otimes}(r,r){\circ}(g,r) \nonumber\\
&{\Rightarrow}(g,g){\circ}(r,r){\circ}(g,r){\otimes}(g,r){\Rightarrow}(g,g){\circ}(r,r){\circ}(g,g){\circ}(r,r) \nonumber
\end{align}

In an irregular bipartite graph, the degree of some vertices could be less than $\Delta$, and a walk on an $(a,b)$ open path may terminate on such a vertex without $a$ or $b$ link. In this case, the missing color can be regarded as the color of a \emph{don't care edge}, denoted as $(\emptyset,\emptyset)$. The color exchange of an $(a,b)$ variable with a don't care edge will always eliminate the variable, denoted as $(a,b){\otimes}(\emptyset,\emptyset)=(a,a)$. The example in Fig.~\ref{KempeWalk}(c) shows that an $(r,b)$ variable was eliminated via color exchange with a don’t care edge.

Despite that Kempe walks can efficiently eliminate most variables, it is obvious that an $(a,b)$ variable in an odd $(a,b)$ cycle cannot be eliminated by a simple Kempe walk. Nevertheless, it is well-known that bipartite graphs do not have odd cycles; therefore, all variables in a bipartite graph can be removed by Kempe walks \cite{complexColor}. Collectively, the on-line scheduling algorithms possess the following features inherent from the properties of complex coloring described in this section.
\begin{enumerate}[{1)}]
\item
Optimality: From K$\ddot{o}$nig’s theorem \cite{graphIntro02}, it is easy to show that the minimum number of colors required for coloring the edges of a bipartite graph equals its maximum degree $\Delta$. Hence, a proper coloring of a graph is optimal if it only uses the minimum number of colors. Suppose a scheduling problem is formulated as a bipartite graph with maximum degree $\Delta$, an initial consistent complex coloring can be easily achieved by $\Delta$ colors, and the color-exchange operations to eliminate variables will not change the color set. Therefore, an optimal $\Delta$-edge coloring of a bipartite graph can always be obtained by Kempe walks.
\item
Parallelizability: In a bipartite graph $G=(X{\cup}Y,E)$, vertices in set $X$ are non-adjacent, and so are vertices in set $Y$. Therefore, simultaneous color exchanges can be performed on vertex sets $X$ and $Y$ alternatively. The next section further elaborates this point.
\item
Rearrangeability: When new calls arrive and existing calls leave the switching system, only partial changes of the existing coloring are needed. Instead of recoloring the entire bipartite graph, any variables generated by edges corresponding to newly arrived calls can be eliminated by Kempe walks.
\end{enumerate}

\section{Parallel Complex Coloring}\label{parallelCColoring}
The efficiency of edge coloring can be substantially improved by parallel color-exchange operations. To guarantee that the number of variables does not increase, complex coloring requires that all color-exchange operations must be effective. It is obvious that this effectiveness condition may be violated by simultaneously performing color-exchange operations on the two end vertices of an edge. For a bipartite graph $G=(X{\cup}Y,E)$, because the vertices in set $X$ are non-adjacent and so are the vertices in $Y$, parallel complex coloring becomes viable according to the following principle.
\newtheorem*{principle1}{Principle of Parallelization}
\begin{principle1}
For a bipartite graph $G=(X{\cup}Y,E)$, simultaneous color exchanges can be performed on vertices in $X$ and $Y$ alternatively.
\end{principle1}

\begin{figure*}[!t]
\centering
\includegraphics[scale=0.90]{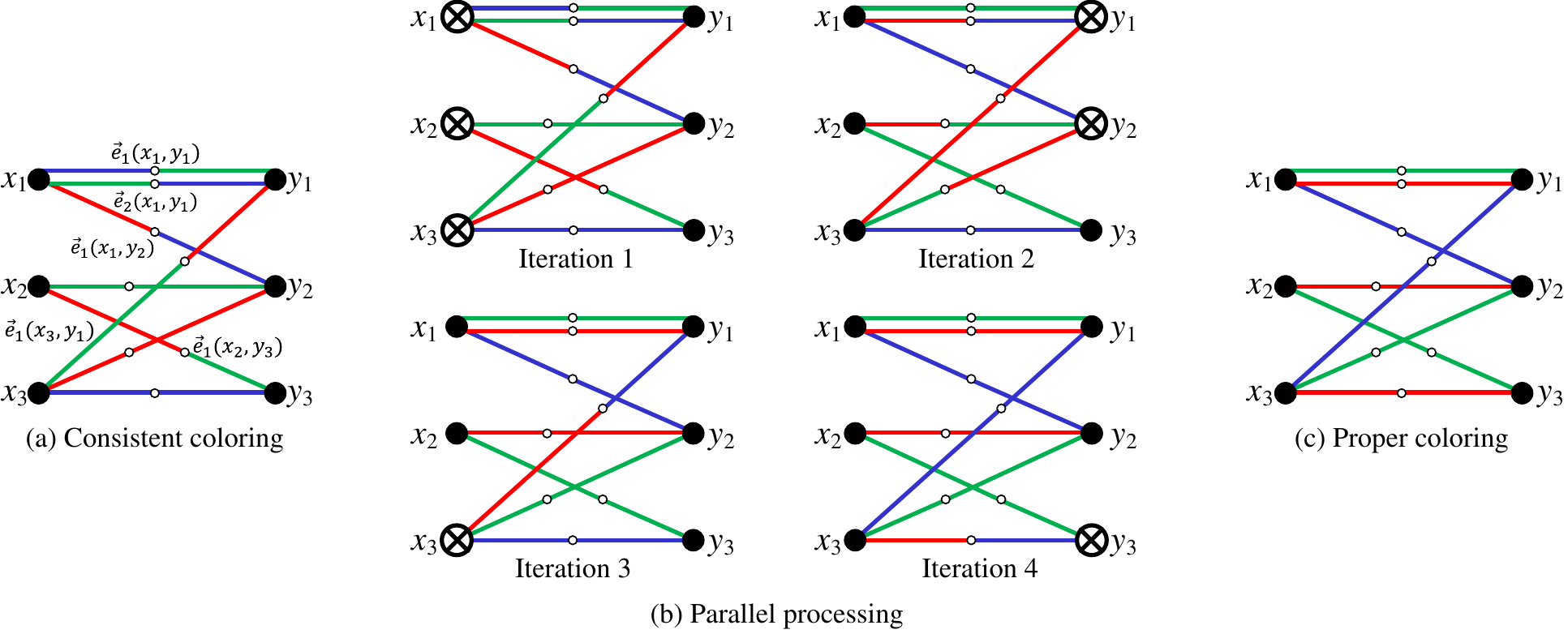}
\caption{Illustration of parallel complex coloring.}
\label{parallelCC}
\end{figure*}

This principle ensures that parallel complex coloring of a bipartite graph is effective. Specifically, the vertices in $X$ execute color-exchange operations in parallel in the first iteration, and color exchanges only performed on the vertices in $Y$ in the next iteration, and so on. Fig.~\ref{parallelCC} demonstrates an example of the parallel processing procedure, starting with the initial consistent coloring shown in Fig.~\ref{parallelCC}(a). In the first iteration, as Fig.~\ref{parallelCC}(b) shows, simultaneous color-exchange operations are performed on vertices in $X$, and then in the next iteration on vertices in $Y$, and so on. When all variables are eliminated, Fig.~\ref{parallelCC}(c) displays the final proper coloring of graph $G$.

In spite of its high efficiency of variable eliminations, parallel complex coloring may introduce deadlock problems when few variables walk in some loops indefinitely, which is the next issue to be addressed in this section. To cope with deadlock variables, we discuss the stopping rules to halt parallel processing in subsection \ref{stopRule}, in which we show that the halting time of parallel processing is on the order of $O(\log⁡|V|)$, where $V=X{\cup}Y$.

\subsection{Deadlock Variables}\label{deadlockIntro}
\begin{figure*}[!t]
\centering
\includegraphics[scale=0.83]{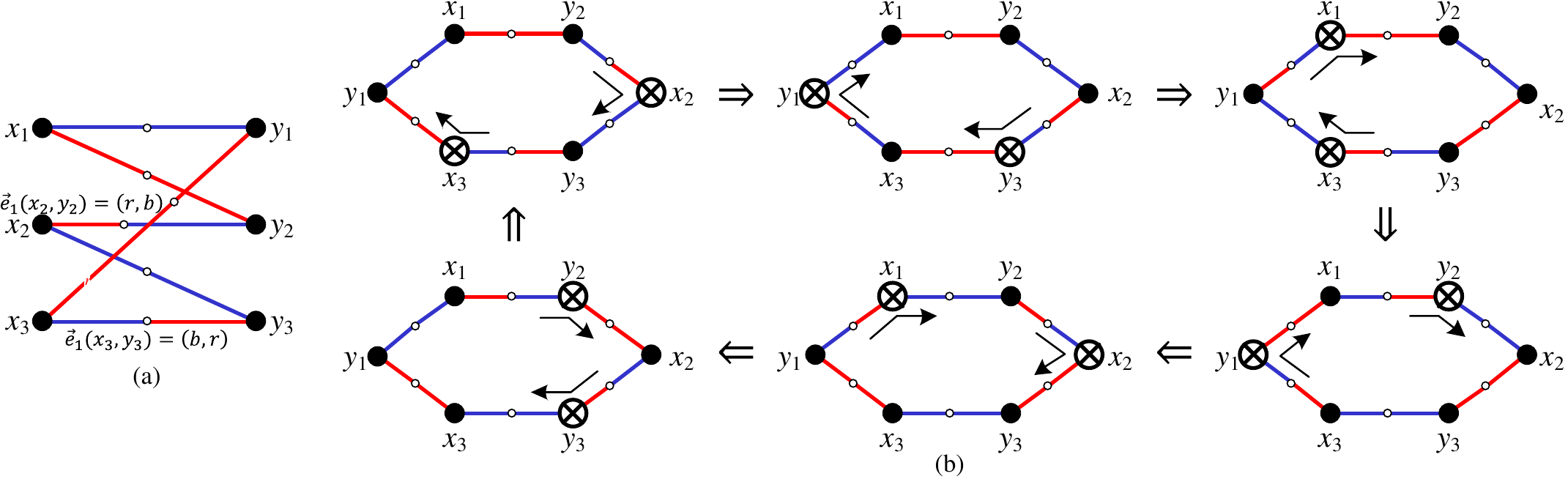}
\caption{Illustration of infinite synchronous walks of deadlock variables.}
\label{deadlock}
\end{figure*}

The parallel processing of complex coloring may introduce deadlock variables, which are trapped in various kinds of infinite loops. The simplest case could be two $(a,b)$ variables synchronously moving in the same direction in a $(a,b)$ cycle. For example, as Fig.~\ref{deadlock}(a) shows, the two variables, $\vec{e}_1(x_2,y_2)=(r,b)$ and $\vec{e}_1(x_3,y_3)=(b,r)$, are blocked in an isolated Kempe cycle. The parallel processing of complex coloring will synchronously move them forward in the same direction such that they will never be eliminated. Fig.~\ref{deadlock}(b) demonstrates the synchronous Kempe walk of these two $(r,b)$ variables in an $(r,b)$ cycle.

\begin{figure}[h]
\centering
\includegraphics[scale=0.93]{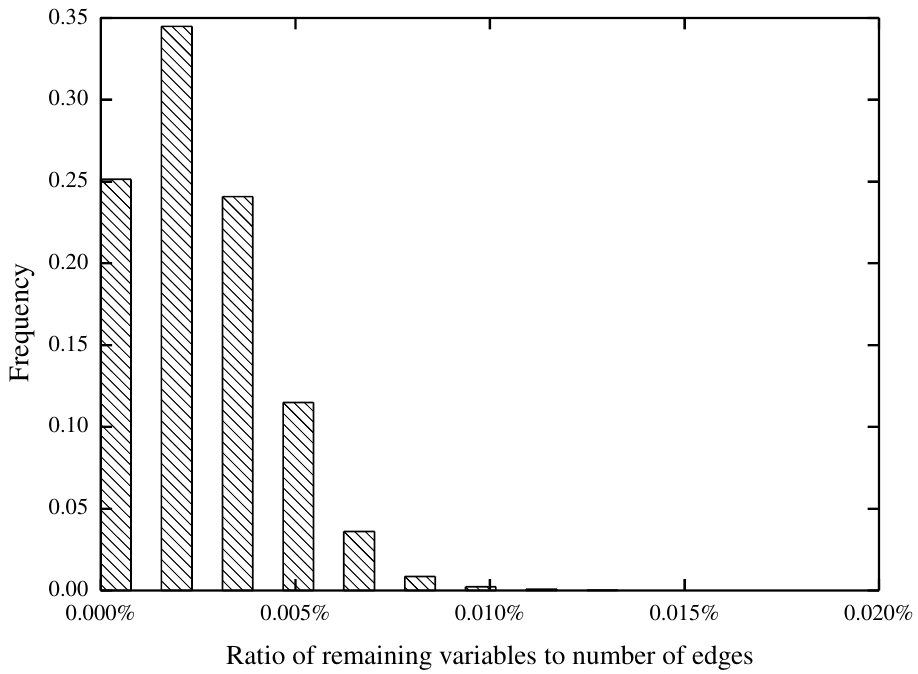}
\caption{Distribution of remaining variable density after 4096 parallel iterations with $|V|=128$ and $\Delta=2000$.}
\label{deadlockPercentage}
\end{figure}

There are numerous topological configurations that may lead to deadlocks because two-colored subgraphs are dynamically changing during the course of the variable elimination process. Fortunately, deadlock loops can be easily broken by variables with overlapped colors; thus, they only occur when the variables are sparse and scattered in relatively isolated two-colored subgraphs. Typically, the density of variables in the graph is very small when deadlocks persist. To confirm this point, we carried out 3000 simulation experiments, where $|V|=128$ and $\Delta=2000$ and we ran at most 4096 iterations for each experiment. The result depicted in Fig.~\ref{deadlockPercentage} shows that the remaining variables versus total number of edges is consistently smaller than 0.01\%. Even though we cannot be sure that all remaining variables are involved in deadlocks, the result clearly indicates that the percentage of deadlock variables is tiny. In the next subsection, we discuss the halting time of parallel complex coloring to prevent aimless moving of variables in the face of deadlocks. 

\subsection{Stopping Rule}\label{stopRule}
We can roughly estimate the halting time of parallel complex coloring from the time complexity of sequential processing. The study in \cite{complexColor} reveals that sequential processing of complex coloring of a bipartite graph, which eliminates variables one at a time, requires $O(|E|\log|V|)$ operations. In parallel processing, if we amortize the operations over all the vertices, then the number of operations implemented by each vertex should be on the order of $O({\Delta}\log⁡|V|)$, which is $1/|V|$ of $O(|E|\log|V|)$. Since each vertex can execute at most $\Delta$ color-exchange operations in each parallel iteration, the expected number of parallel iterations is again on the order of $O(\log⁡|V|)$.

The variable elimination process in complex coloring requires the moving of variables on Kempe paths to meet with other variables. Presumably, the number of iterations of parallel complex coloring is proportional to the average length of a Kempe path. We need the following lemma that has been proved in \cite{pathLength}:
\newtheorem{Lemma1}{Lemma}
\begin{Lemma1}
The average path length in a random graph $G=(V,E)$ is on the order of $O(\log|V|)$.
\label{pathLength} 
\end{Lemma1}

It is known that any acyclic path is a subgraph of a spanning tree \cite{graphIntro02}. An intuitive explanation of the above statement is that the average distance from the root of a spanning tree to any leaf is on the order of $O(\log⁡|V|)$. The statement of Lemma~\ref{pathLength} is consistent with our previous estimate from sequential complex coloring that the number of effective iterations of parallel complex coloring should be on the order of $O(\log⁡|V|)$. We analyze this relationship in more details next. In the analysis and simulation studies to determine the proper halting time of the parallel processing of complex coloring, we need the following definitions to facilitate our discussions.
\newtheorem{def4}[def1]{Definition}
\begin{def4}
We adopt following notations in parallel complex coloring:
\begin{enumerate}[{1)}]
\item
The \textbf{variable density} is the ratio of the number of variables to the number of edges $|E|$; variable density after $t$ iterations is denoted by $R(t)$.
\item
The \textbf{variable elimination rate} is the ratio of the number of eliminated variables to the total number of variables in an iteration; the elimination rate of $t^{th}$ iteration is denoted by $\alpha(t)$.
\item
The \textbf{hitting time} is the expected number of iterations needed for a variable to hit another variable; the hitting time of $t^{th}$ iteration is denoted by $h(t)$. That is, the hitting time is the average number of moves needed to eliminate a variable, and it is inversely proportional to the variable elimination rate.
\end{enumerate}
\end{def4}

The following theorem states the relationship among these variables during the elimination process. We show that the expected number of iterations to halt parallel processing of complex coloring is on the order of $O(\log⁡|V|)$, under the assumption that the variable elimination rate $\alpha(t)$ is a constant.
\newtheorem{Theorem1}{Theorem}
\begin{Theorem1}
If the elimination rate $\alpha(t)$ is a constant with respect to the number of iterations $t$, then the total number of iterations needed for parallel complex coloring to achieve a fixed given variable density $\epsilon$ is on the order of $O(\log⁡|V|)$.
\label{totalIterations} 
\end{Theorem1}
\begin{IEEEproof}
Suppose that the variable elimination rate is a constant $\alpha$ with respect to $t$, and then the expected hitting time is also a constant, which is inversely proportional to $\alpha$ as follows:
\begin{equation}
\alpha=\frac{a}{h},
\end{equation}
where $a$ is a positive constant. The number of remaining variables at the end of $t^{th}$ iteration is $|E|R(t)$, and the number of variables that will be eliminated in $(t+1)^{th}$ iteration is $|E|R(t){\times}\alpha$. Thus, we have
\begin{equation}
|E|R(t)-|E|R(t+1)=\alpha|E|R(t),
\end{equation}
which yields
\begin{equation}
R(t)=(1-\alpha)^tR(0),
\end{equation}
where $0{\leq}R(0){\leq}1$ is the initial variable density. For achieving a variable density $\epsilon$, the required number of iterations $T$ is given by
\begin{equation}
(1-\alpha)^TR(0)=\epsilon.
\end{equation}
For $\alpha{\ll}1$, we can obtain
\begin{equation}
T{\approx}\frac{\ln{\epsilon}-\ln{R(0)}}{-\alpha}=\frac{1}{\alpha}\ln\frac{R(0)}{⁡\epsilon}=\frac{h}{a}\ln\frac{R(0)}{⁡\epsilon}.
\end{equation}
According to Lemma~\ref{pathLength}, hitting time $h$ is on the order of $O(\log⁡|V|)$. Therefore, the number of iterations needed to achieve the specific variable density $\epsilon$ is also on the order of $O(\log⁡|V|)$.
\end{IEEEproof}

\begin{figure}[h]
\centering
\includegraphics[scale=0.93]{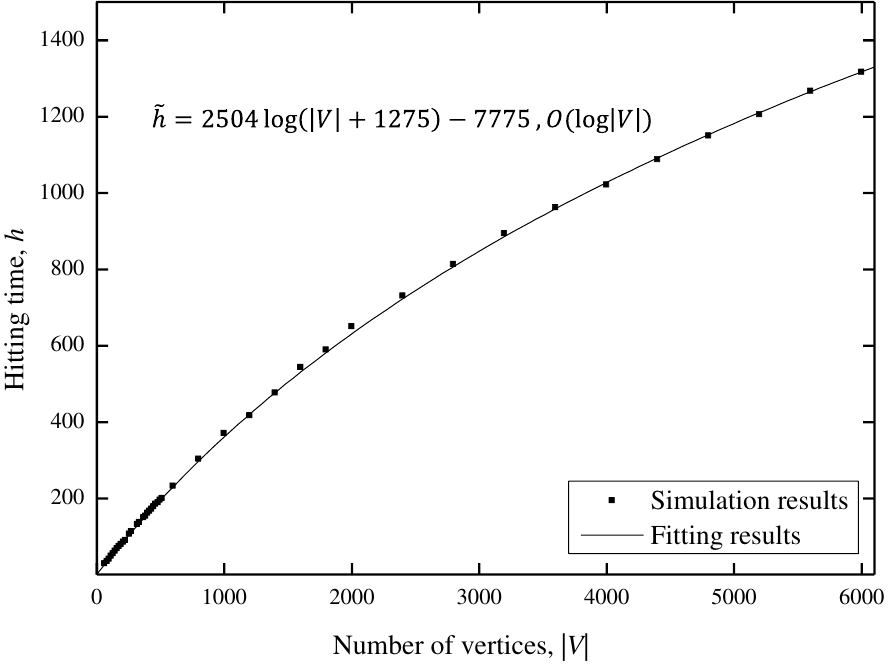}
\caption{The hitting time $h$ as a function of the number of vertices $|V|$.}
\label{hittingTime}
\end{figure}

In Fig.~\ref{hittingTime}, the simulation result further confirms the statement of Theorem~\ref{totalIterations} that hitting time $h$ is on the order of $O(\log⁡|V|)$. Thus, the number of iterations needed to achieve the specific variable density $\epsilon$ is $O(\log⁡|V|)$.

\begin{figure}[h]
\centering
\includegraphics[scale=0.93]{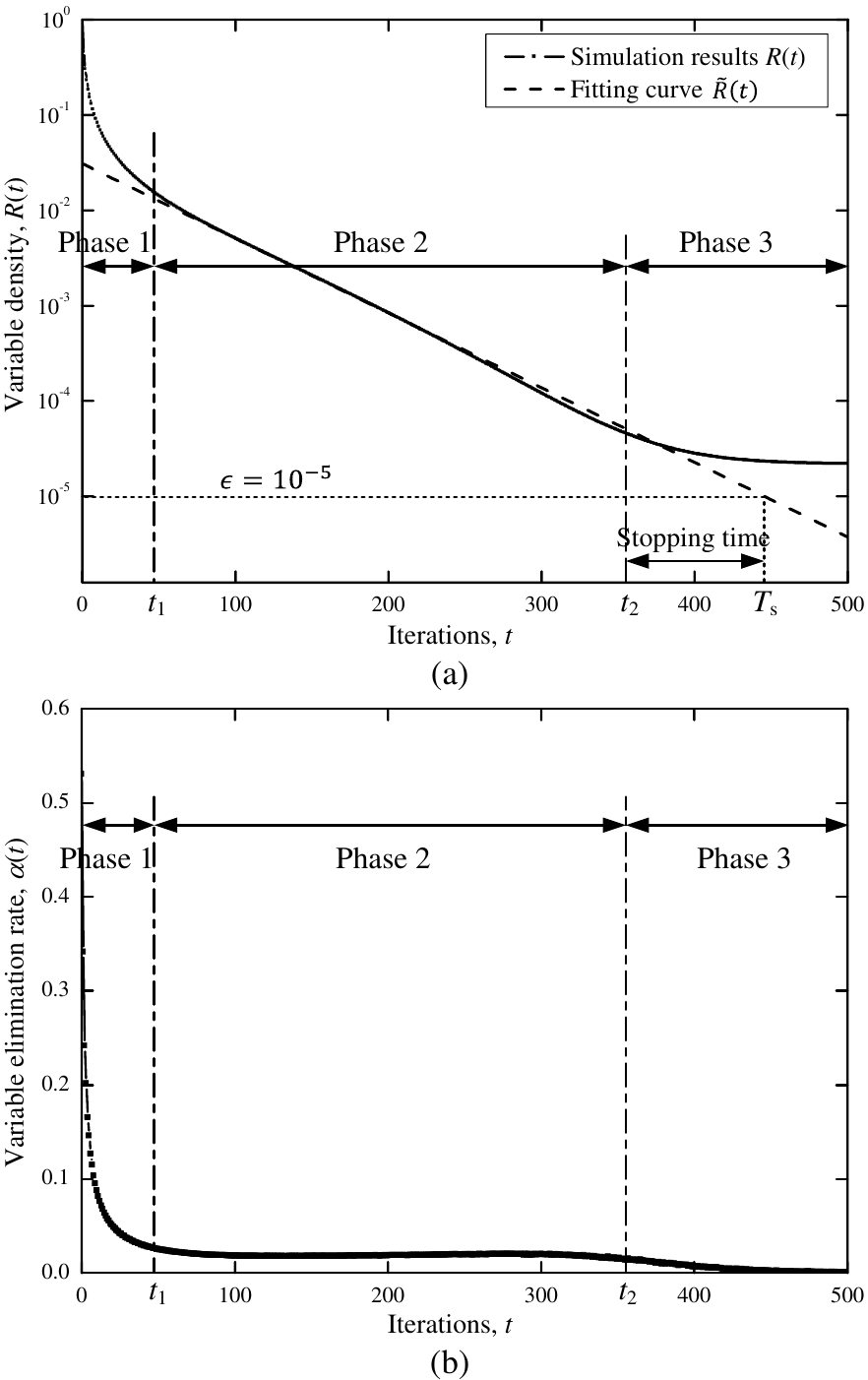}
\caption{The three phases of parallel complex coloring of a bipartite graph with $|V|=128$ and $\Delta=2000$.}
\label{Rt_alpha}
\end{figure}

In reality, however, the variable elimination rate $\alpha(t)$ is not a uniform constant with respect to the number of iterations $t$ . For a randomly generated bipartite graph with $|V|=128$ vertices and maximum degree $\Delta=2000$, the simulation results displayed in Fig.~\ref{Rt_alpha}(a) and (b) show that the parallel processing of complex coloring actually goes through the following three phases:
\begin{enumerate}[{1)}]
\item
In the first phase, the initial variable elimination rate $\alpha(t)$ is very large but drops rapidly; therefore the remaining variable density $R(t)$ decreases very fast.
\item
In the second phase, the variable elimination rate $\alpha(t)$ is stabilized and becomes a constant, and the variable density $R(t)$ declines linearly with the same slope. 
\item
In the third phase, the variable elimination rate $\alpha(t)$ is almost equal to 0, and the variable density $R(t)$ remains unchanged.
\end{enumerate}

The above three phases can be perfectly explained by the characteristics of parallel complex coloring. Intuitively, variables are more likely to be eliminated when they are close together; therefore, with high variable density $R(t)$, the variable elimination rate $\alpha(t)$ is very large and the variable density decreases very fast in phase 1. As the variable density $R(t)$ decreases, the hitting time increases and thus the elimination rate slows down in phase 2. When almost all variables have been eliminated, the elimination rate decreases to zero and the variable density remains unchanged because most remaining variables are likely blocked in deadlock loops. This is the behavior of phase 3 that we have observed from simulation results. Formally, we define the three phases according to the variable density $R(t)$ and variable elimination rate $\alpha(t)$ as follows.
\newtheorem{def5}[def1]{Definition}
\begin{def5}
The three phases of parallel complex coloring are characterized by the changes of variable density $R(t)$ and variable elimination rate $\alpha(t)$ as follows:
\begin{enumerate}[{1)}]
\item
Initial phase: ${\bigtriangleup}R(t)<0$,${\bigtriangleup}\alpha(t)<0$. Variable density decreases drastically and so does the elimination rate.
\item
Steady phase: ${\bigtriangleup}R(t)<0$,${\bigtriangleup}\alpha(t)=0$. Variable density decreases, but the elimination rate is a constant.
\item
Deadlock phase: ${\bigtriangleup}R(t)=0$,${\bigtriangleup}\alpha(t)=0$. Variable density approaches zero, and no more variable can be eliminated; thus the elimination rate is zero.
\end{enumerate}
\end{def5}

The halting time of parallel processing should be predetermined to avoid aimless operations in phase 3, meaning the parallel algorithm should be stopped at the end of phase 2. In particular, we need to determine the time $t_1$, when the elimination process enters the steady phase from the initial phase, and the time $t_2$, when the steady phase ends and the deadlock phase begins. In practice, we use the following more relaxed conditions to each epoch:
\begin{enumerate}[{1)}]
\item
Initial phase: ${\bigtriangleup}R(t)<0$,${\bigtriangleup}\alpha(t)<0$.
\item
Steady phase: ${\bigtriangleup}R(t)<0$,${\bigtriangleup}\alpha(t)<10^{-3}$.
\item
Deadlock phase: ${\bigtriangleup}R(t)<10^{-6}$,${\bigtriangleup}\alpha(t)<10^{-3}$.
\end{enumerate}

The three phases demonstrated in Fig.~\ref{Rt_alpha}(a) and (b) are determined by the above criteria. In the initial phase, the variable density $R(t)$ decreases very fast from 1 to $10^{-3}$. In the steady phase, the elimination rate $\alpha(t)$ is a constant and $R(t)$ declines from $10^{-2}$ to $10^{-4}$ almost with the same slope. In the deadlock phase, the elimination rate $\alpha(t)$ is close to $0$ and the variable density $R(t)$ remains unchanged, mainly due to deadlocks. 

As for the selection of the stopping rule, only the beginning and the ending of the steady phase are of interest, since the initial phase is too short, and in practice the deadlock phase will be treated in a different way, which will be discussed in the next section. In the steady phase, the elimination rate $\alpha(t)$ converges to a constant $\alpha$ and the hitting time $h(t)$ becomes a constant $h$ with respect to $t$, which is conformable to the assumption of Theorem~\ref{totalIterations}. Therefore, following the proof of Theorem~\ref{totalIterations}, the variable density in the steady phase can be expressed as follows:
\begin{equation}
R(t)=(1-\alpha)^{t-t_1}R(t_1),
\label{steadyRt}
\end{equation}
where $t{\in}[t_1,t_2]$. In Fig.~\ref{Rt_alpha}(a), the dashed line, denoted as $\widetilde{R}(t)$, is plotted according to (\ref{steadyRt}). Since the line $\widetilde{R}(t)$ starts with the initial value $R(t_1)$ at $t=t_1$ with the constant slop $\alpha$, $\widetilde{R}(0)<R(0)$ and the variable density $R(t)$ is lower bounded by $\widetilde{R}(t)$ throughout all $t$. For a given variable density $\epsilon$, the upper bound of the stopping time is given by:
\begin{equation}
T_s=-\frac{\ln{\epsilon}-\ln{R(t_1)}}{\alpha}+t_1=\frac{h}{a}\ln{\frac{R(t_1)}{\epsilon}+t_1},
\label{Ts}
\end{equation}
where $a$ is a constant. Since the steady phase ends at time $t_2$, the stopping time can be chosen from the interval $[t_2,T_s]$. We summarize our discussions on the stopping rule in the following principle.
\newtheorem*{principle2}{Principle of Stopping Rule}
\begin{principle2}
To achieve a given remaining variable density $\epsilon$, the parallel complex coloring of a bipartite graph $G=(V,E)$ should halt after a $\log⁡(|V|+b)+c$ iterations, where $a$,$b$ and $c$ are application-specific parameters.
\end{principle2}

There is a trade-off between the stopping time and the number of remaining variables. If the running time is a major concern, then the stopping time should be chosen close to $t_2$. Otherwise, if a small number of remaining variables is of interest in the scheduling application, then the stopping time can be chosen closer to the upper bound $T_s$. The application of the Principle of Stopping Rule is further discussed in the next section.

\section{Parallel Scheduling Algorithms}\label{schedulingAlgorithms}
This section describes the design of an on-line scheduling algorithm for packet switching. As Section~\ref{parallelCColoring} demonstrates, parallel Kempe walks may not be able to completely eliminate remaining variables in the deadlock phase of parallel complex coloring algorithm. We discuss a strategy in dealing with remaining deadlock variables in this section.  

There is a lot of concern over the time complexity for on-line scheduling of packet switching. If the running time of scheduling is a stringent requirement, keeping down deadlock variables and recoloring them in next time frame is the only viable strategy. As Fig.~\ref{deadlockPercentage} shows, the wasted bandwidth due to recoloring remaining variables is almost negligible. Thus, the development of an ultra-fast on-line parallel scheduling algorithm by complex coloring to achieve a high throughput for frame-based packet switching is feasible. 

In this section, we first describe the parallel algorithm of complex coloring for bipartite graphs, and then discuss the application of this algorithm to on-line scheduling for frame-based packet switching.

\subsection{Parallel Complex Coloring Algorithm}\label{ParallelCCAlgorithm}
Based on previously proposed Principle of Parallelization and Principle of Stopping Rule, we develop a parallel iterative algorithm of complex coloring for a bipartite graph $G=(X{\cup}Y,E)$. In each iteration, the color exchanges are operating on the vertices in set $X$ and then on vertices in set $Y$. The algorithm repeats this iterative procedure until all variables are eliminated or takes a time out when the stopping rule is satisfied. We now present the algorithm \emph{Parallel Complex Coloring} as Algorithm~\ref{parallelCCAlg} in this subsection.
\begin{algorithm}[!t]
\caption{Parallel Complex Coloring}
\label{parallelCCAlg}
\begin{algorithmic}[1]
\REQUIRE 
A consistently colored bipartite graph $G=(X{\cup}Y,E)$, color set $C$, $L(x_i)$ (or $L(y_i)$): the list of variables incident to vertex $x_i$ (or $y_i$), stopping time $T{\in}[t_2,T_s]$. 
\ENSURE 
A properly colored graph $G$, or a consistently colored graph $G$ with few remaining variables.
\STATE
For each vertex $x_i$ in $X$, update $L(x_i)$ and execute color-exchange operations in parallel: If $L(x_i)$ is nonempty, then for each variable $\vec{e}=(a,b)$ in $L(x_i)$, find its adjacent link of $(a,b)$ path and do the color-exchange operation.
\STATE
For each vertex $y_i$ in $Y$, update $L(y_i)$ and execute color-exchange operations in parallel: If $L(y_i)$ is nonempty, then for each variable $\vec{e}=(a,b)$ in $L(y_i)$, find its adjacent link of $(a,b)$ path and execute the color-exchange operation.
\STATE
Repeat Steps 1-2, until one of the following two conditions is satisfied:
\begin{enumerate}[{C1:}]
\item
All variables in graph $G$ have been eliminated.
\item
The number of iterations reaches the predetermined stopping time $T$.
\end{enumerate}
\end{algorithmic}
\end{algorithm}

The parallel complex coloring algorithm will return a properly colored graph $G$ with a high probability in $O(\log⁡|V|)$ iterations. Since there could be at most $\Delta$ variables incident to a vertex in each iteration, the time complexity of the Parallel Complex Coloring algorithm is on the order of $O(\Delta\log⁡|V|)$. The leftover variables after the parallel algorithm stops are simply kept down in the next frame because they could be trapped in deadlock loops in the current time frame.

\subsection{On-line Scheduling for Frame-based Packet Switching}\label{onlineAlgorithm}
We consider an $N{\times}N$ input-queued switch and define the following notations to facilitate our discussion. Suppose the time is divided into frames of size $f$, a bipartite graph $G_k=(X{\cup}Y,E)$ is constructed for packets that arrive in the $k^{th}$ time frame. Vertex $x_i$ in $X$ and vertex $y_j$ in $Y$ represent input $i$ and output $j$, respectively, and edge $e_t(x_i,y_j){\in}E$ stands for the $t^{th}$ packet in the buffer of input $i$ destined for output $j$, where $i,j=1,2,\dots,N$. Let $\Delta$ be the maximum degree of graph $G$ and $C=\{c_1,c_2,\dots,c_{\Delta}\}$ be the set of colors, each of which corresponds to a timeslot in the next time frame. For example, suppose edge $e_t(x_i,y_j)$ is colored by color $c_{\gamma}$, for some $\gamma=1,2,\dots,\Delta$, then the corresponding packet will be sent to output $j$ in the γth timeslot of the next time frame.

Furthermore, for two consecutive bipartite graphs $G_{k-1}$ and $G_k$, if the difference of their edge sets is small, which is common in core switches \cite{frameGame}, then the graph $G_k$ in $k^{th}$ time frame can be obtained by slightly modifying the edge set of $G_{k-1}$. According to the rearrangeability property of complex coloring described in Section~\ref{preliminary}, instead of recoloring all edges of $G_k$, a proper coloring of $G_k$ can be obtained by only consistently coloring the newly created edges in $G_k$ and then eliminating the newly generated variables by Kempe walks. Therefore, each input and output can accomplish the graph initialization independently in each time frame with the new packet arrivals and previous coloring information. In Appendix~\ref{appendix1}, we describe the implementation of \emph{Graph Initialization} process in more details. 

Frame size $f$ is selected under the assumption that the following traffic load constraints are observed: $\lambda_i=\sum_{j}\lambda_{ij}<1$ and $\lambda_j=\sum_{i}\lambda_{ij}<1$, where $\lambda_{ij}$ is the average traffic rate from input $i$ to output $j$ in a timeslot. Ideally, packets accumulated in the current time frame should all be sent out in the next time frame, which means $f$ colors should be sufficient to properly color the bipartite graph $G_k$ in any time frame $k$. We know from the condition of edge coloring of bipartite graph that a proper coloring with $f$ colors can always be obtained if the maximum degree $\Delta{\leq}f$ \cite{graphIntro02}. This condition can be easily satisfied when the arrival rate is low. However, when the traffic load is heavy, the maximum degree $\Delta$ may exceed frame size $f$ due to the randomness of output addresses of input packets. If $\Delta>f$, packets that cannot be scheduled are deferred. Thus, to enhance the throughput, it is necessary to select a proper frame size $f$ to ensure that $\Delta{\approx}f$ with high probability. In Appendix~\ref{appendix3}, we show that, if the frame size is sufficiently large, say $f{\geq}O(\log⁡N)$, the number of packets in the buffer of input $i$ destined to output $j$ in a time frame will converge to the average number $f\lambda_{ij}$, such that the maximum degree $\Delta$ will be close to frame size $f$ with high probability, that is, $\Delta{\approx}\sum_{i}f\lambda_{ij}<f$ and $\Delta{\approx}\sum_{j}f\lambda_{ij}<f$.

The frame-based scheduling algorithm is completely parallel and distributed. The on-line algorithm starts with a consistently colored bipartite graph for each frame of packets, and then iteratively calculates a set of matchings by variable eliminations. The parallel algorithm stops after $T$ iterations, which is predetermined based on the remaining variable density $\epsilon$ required by the scheduler. The remaining variables, presumably most of them are deadlock variables, are ignored in the current time frame and kept down in the next time frame. Let $M_{i,j}$ denote the set of edges incident to vertex $x_i{\in}X$ and vertex $y_j{\in}Y$ that will be recolored in next time frame. This information will be stored in each input line card and processed in the next frame. The on-line frame-based scheduling algorithm is described as follows.
\begin{enumerate}[{Step 1}]
\item
Initialization: Each input and output implement \emph{Graph Initialization}, according to the frame of packets, the uncolored edges $M_{i,j}$ of the previous time frame, and the previous coloring information in the case of rearranging existing connections.
\item
Perform \emph{Parallel Complex Coloring}, and return the colored bipartite graph with the remaining variables after $T$ iterations.
\item
Carryover Information: For each input $i$, update the set of remaining variables $M_{i,j}$ from the result of Step 2.
\end{enumerate}

\begin{figure}[h]
\centering
\includegraphics[scale=1.15]{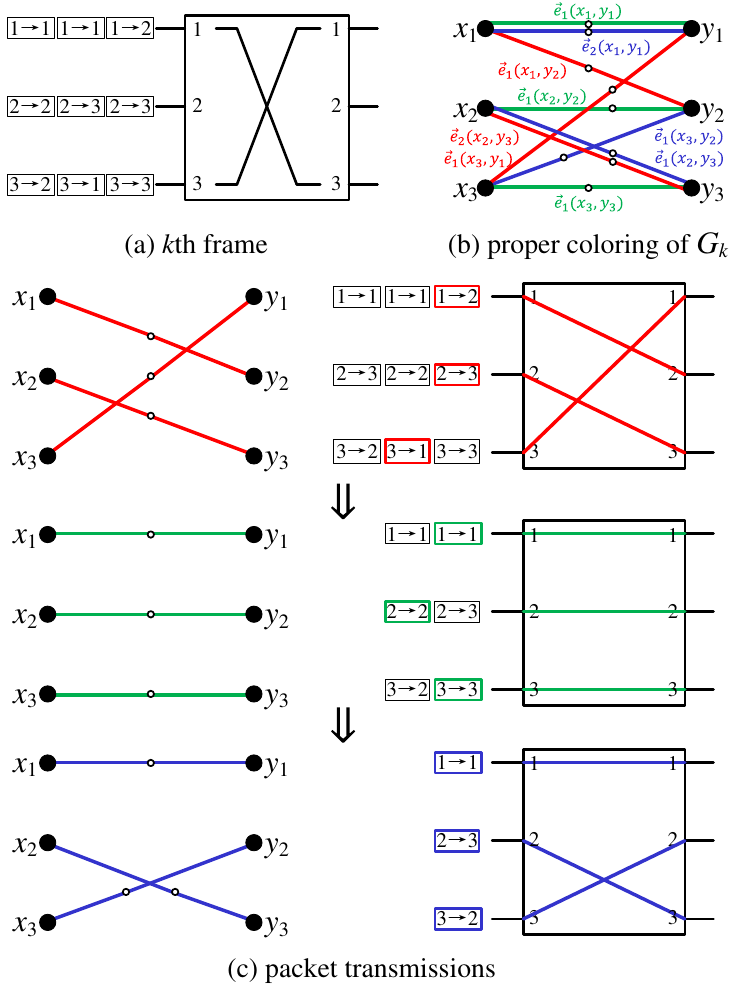}
\caption{Packet transmission according to the coloring of bipartite graph.}
\label{transmission}
\end{figure}

In the colored bipartite graph, edges of the same color constitute a matching that represents the scheduled permutation of a corresponding timeslot. It should be noted that a straightforward application of the coloring graph to schedule the packets may introduce out-of-sequence packets. This occurs when an edge corresponds to an earlier arriving packet but is colored by a higher indexed color. For example, as Fig.~\ref{transmission}(a) shows, the input packets are ordered by their arrival times, and the indices of the colors in the set $C=\{c_1,c_2,c_3\}=\{r,g,b\}$ represent the corresponding timeslots in a frame. As Fig.~\ref{transmission}(b) illustrates, the two edges $e_1(x_2,y_3)$ and $e_2(x_2,y_3)$ between input $x_2$ and output $y_3$ are respectively colored by $c_3$ and $c_1$, which means the first packet $e_1(x_2,y_3)$ will be sent in the third timeslot but the second packet $e_2(x_2,y_3)$ in the first timeslot. If these two packets belong to the same session, they will be received out of their original delivering sequence.

The out-of-sequence problem is caused by the inconsistency between the order of arrivals and the order of the color indices. This problem can be easily resolved by sorting the colors of those edges between any vertex pair according to their arrival times, and then each input can determine the packet sending sequence in a time frame. After sorting the colors, as Fig.~\ref{transmission}(c) shows, input 2 sends the first packet in the first timeslot and the second packet in the third timeslot, which are in the same order of their arrival times.

\section{Performance of Scheduling Algorithms}\label{performance}
This section evaluates performances of scheduling algorithms. There are three aspects, throughput, delay, and complexity, concerning the performance of an on-line scheduling algorithm. In subsection~\ref{delayThroughput}, delay and throughput performances are presented under different traffic patterns. In subsection~\ref{complexity}, we present complexity analyses of on-line algorithms.

\subsection{Delay and Throughput}\label{delayThroughput}
In this subsection, simulation results of on-line frame-based scheduling algorithm are presented in comparison with the well-known \emph{i}SLIP algorithm. In the simulation, we considered both uniform and non-uniform traffic. The later includes diagonal hotspot traffic \cite{APSARA03}, and log-diagonal traffic \cite{frameMatching}. We show that our algorithm can achieve a high throughput as well as an acceptable average delay under uniform input traffic, and its performance is highly robust under both non-uniform traffic models.

\subsubsection{Uniform Traffic}\label{uniform}
In our simulation, we assume that the arrival process at each input of an input-queued switch is a Bernoulli process with rate $\lambda$, and the destination of each packet is uniformly distributed over all outputs, which means $\lambda_{ij}=\lambda/N,{\forall}i,j$. 

\begin{figure}[h]
\centering
\includegraphics[scale=0.93]{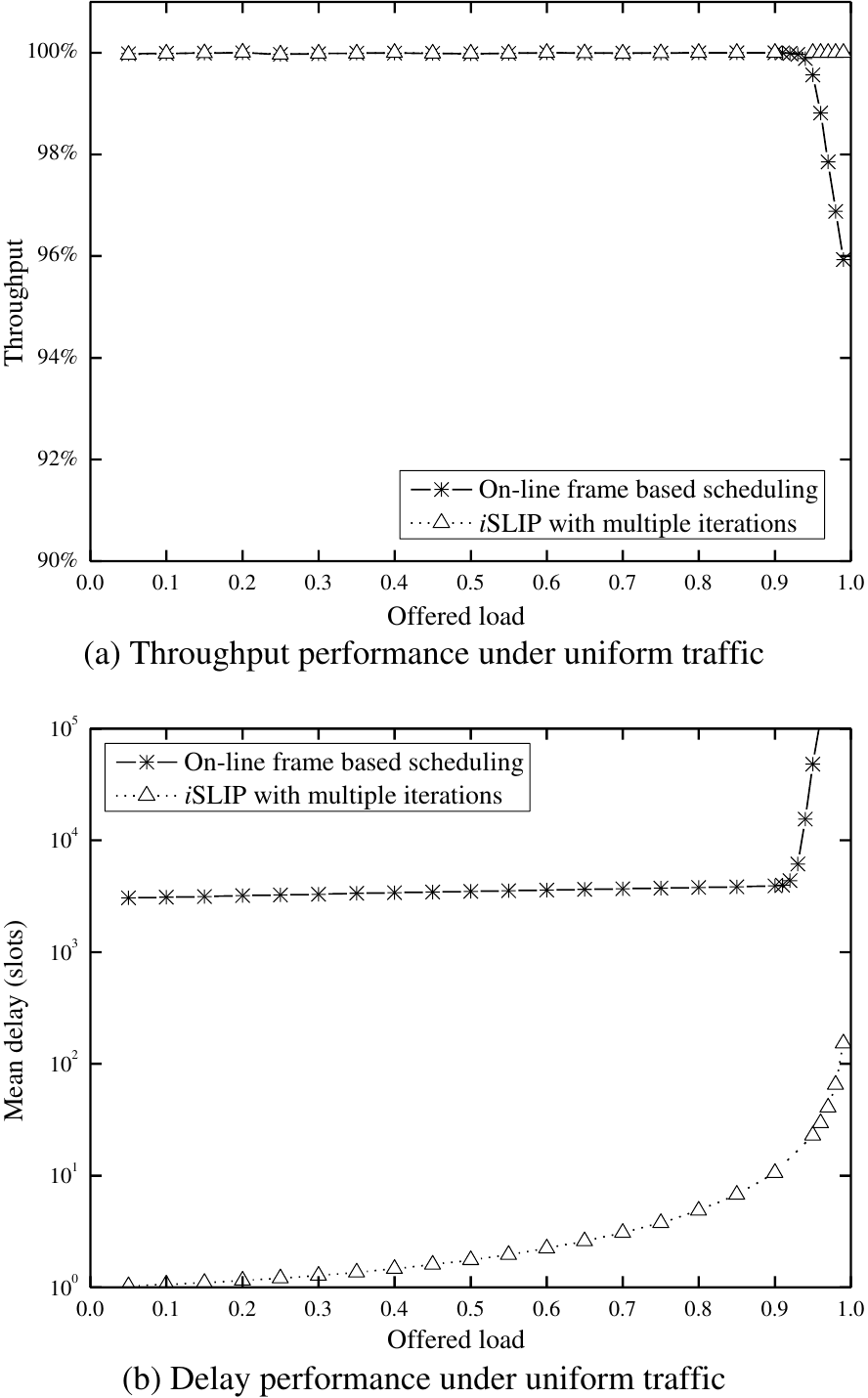}
\caption{Performance comparison of the scheduling algorithms under the uniform traffic.}
\label{performanceUniform}
\end{figure}

The performance of our algorithm under uniform input traffic is plotted in Fig.~\ref{performanceUniform}, where $N=64$ and $f=2000$. As we mentioned in Section~\ref{onlineAlgorithm} and explained in Appendix~\ref{appendix3}, the maximum degree $\Delta$ of the bipartite graph in a frame is typically larger than $f$ when each input is fully loaded. If $\Delta>f$ in a frame, then not all accumulated packets can be cleared out from input buffers, simply because there are not enough timeslots, maximum is $f$, to schedule $\Delta$ packets within this time frame. It follows that the system becomes unstable and the delay goes to infinity, as shown in Fig.~\ref{performanceUniform}, when the traffic load is larger than 0.95. On the other hand, when the traffic load is below 0.9, the maximum degree $\Delta$ is almost always less than $f$. In this case, a close to 100\% throughput and a bounded average delay can be achieved, which coincides with the simulation results in Fig.~\ref{performanceUniform}.

Under uniform input traffic, as the comparison in Fig.~\ref{performanceUniform} shows, the \emph{i}SLIP outperforms our on-line scheduling algorithm in both throughput and delay. The \emph{i}SLIP can achieve 100\% throughput under a full load since it approximately performs as a time-division multiplexing scheme \cite{iSLIP}. Moreover, the \emph{i}SLIP operates in a slot-by-slot manner, while our algorithm is frame-based. Compared with the \emph{i}SLIP, our algorithm will suffer from a larger packet delay due to the waiting time spent in the accumulating stage and scheduling stage. Nevertheless, as current state-of-the-art technology demonstrates, the timeslot is getting shorter and shorter (e.g., nanoseconds) as the line rate of each port increases up to 100 Gb/s \cite{refOIDA}. Therefore our scheduling algorithm can easily meet the most stringent delay requirement, which is on the order of milliseconds \cite{frameGame}.

\subsubsection{Non-uniform Traffic}\label{non-uniform}
\begin{figure*}[!t]
\centering
\subfigure{
\label{performanceNonuniform-a}
\includegraphics[scale=0.93]{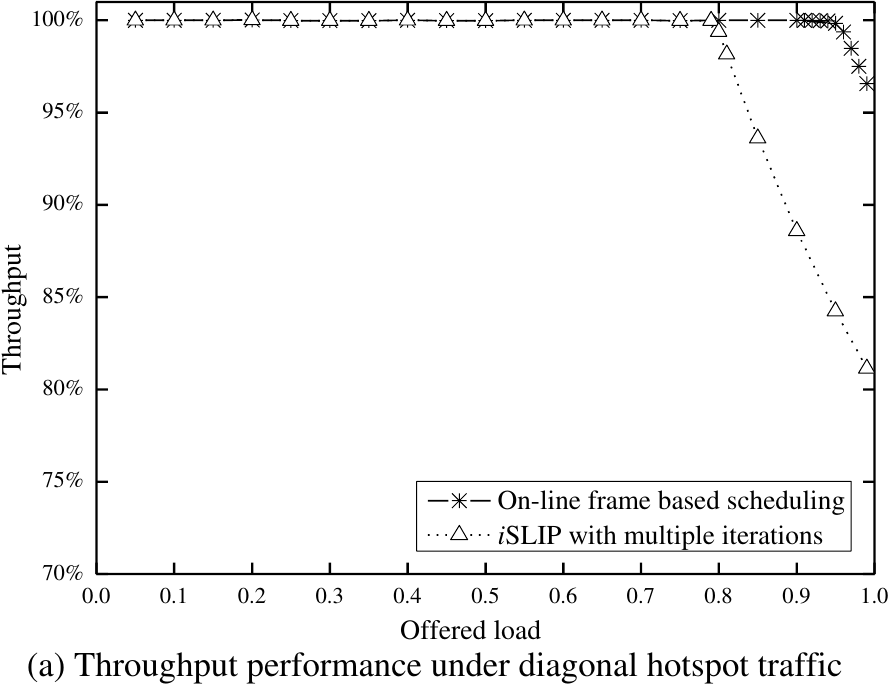}}
\subfigure{
\label{performanceNonuniform-b}
\includegraphics[scale=0.93]{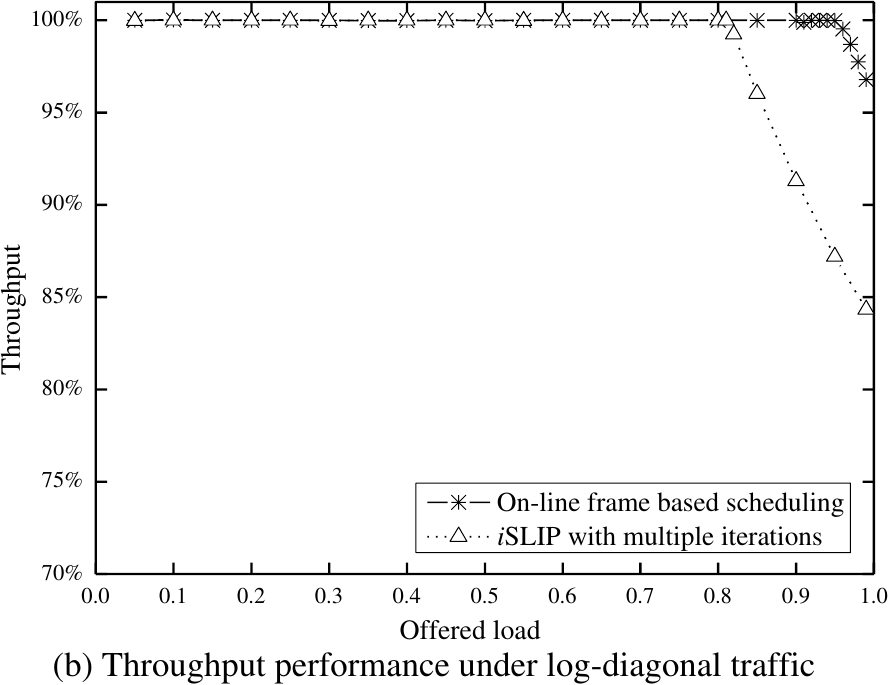}}
\subfigure{
\label{performanceNonuniform-c}
\includegraphics[scale=0.93]{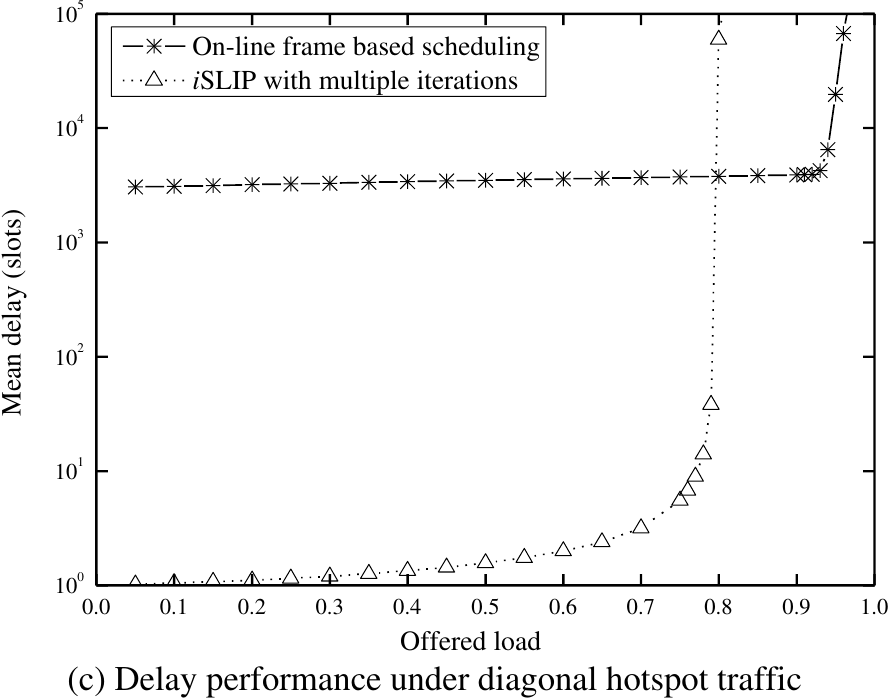}}
\subfigure{
\label{performanceNonuniform-d}
\includegraphics[scale=0.93]{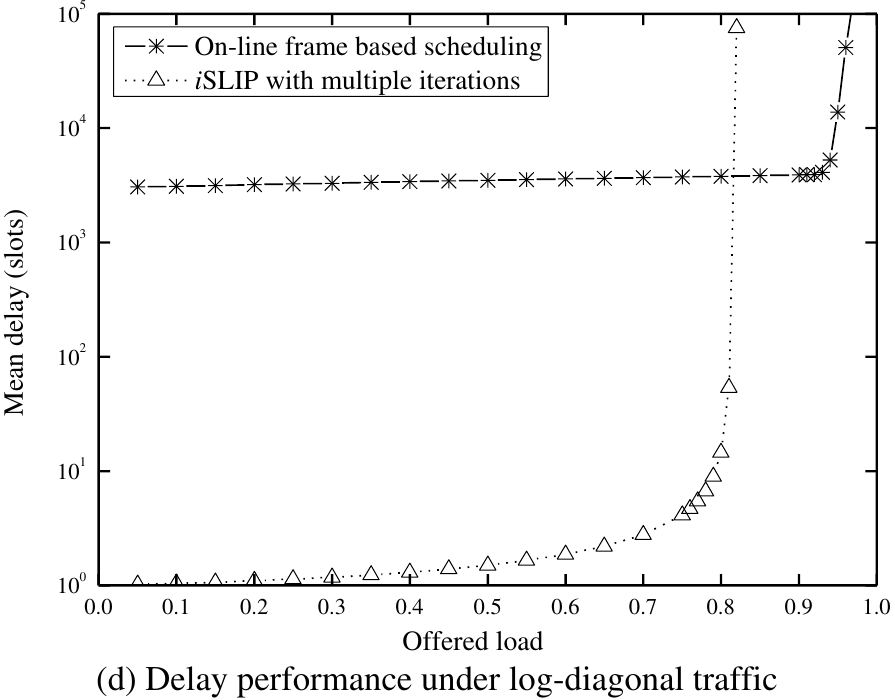}}
\caption{Performance comparison of the scheduling algorithms under the nonuniform traffic.}
\label{performanceNonuniform}
\end{figure*}

Under non-uniform input traffic, the simulation results demonstrate the robustness of our complex-coloring-based scheduling algorithm. Since the efficiency of the variable elimination process becomes much higher when variables are clustered together, the uneven distribution of traffic load does not adversely affect the performance of our scheduling algorithm. In our simulation study, we consider two popular non-uniform traffic models, diagonal hotspot traffic \cite{APSARA03}, and log-diagonal traffic \cite{frameMatching}. They are specifically defined as follows:
\begin{enumerate}[{1)}]
\item
Diagonal hotspot \cite{APSARA03}: $\lambda_{ii}=\frac{\lambda}{2}$ and $\lambda_{ij}=\frac{\lambda}{2(N-1)}$, ${\forall}i,j$.
\item
Log-diagonal \cite{frameMatching}: $\lambda_{ii}=\frac{2^{(N-1-i+j){\bmod}N}}{(2^N-1)}$, ${\forall}i,j$.
\end{enumerate}

The simulation results of the scheduling algorithms displayed in Fig.~\ref{performanceNonuniform}, where $N=64$ and $f=2000$, shows that the performance of our on-line scheduling algorithm under non-uniform traffic is as good as that under uniform traffic. In addition to the high efficiency of eliminating clustered variables by complex coloring, the frame-based algorithm actually serves as a time-division statistical multiplexer for packets within a frame to smooth out the unevenness of the traffic load. 

Meanwhile, the \emph{i}SLIP algorithm becomes unstable under non-uniform traffic when the input traffic load is high. As Fig.~\ref{performanceNonuniform}(a) and (b) shows, when the offered load approaches 1, the \emph{i}SLIP only achieves 81.1\% throughput under diagonal hotspot traffic and 84.3\% under log-diagonal traffic. Furthermore, for both non-uniform input traffic patterns, the average delay of \emph{i}SLIP skyrockets to infinity when the offered load is above 80\%.

\subsection{Complexity}\label{complexity}
\begin{figure}[h]
\centering
\includegraphics[scale=0.93]{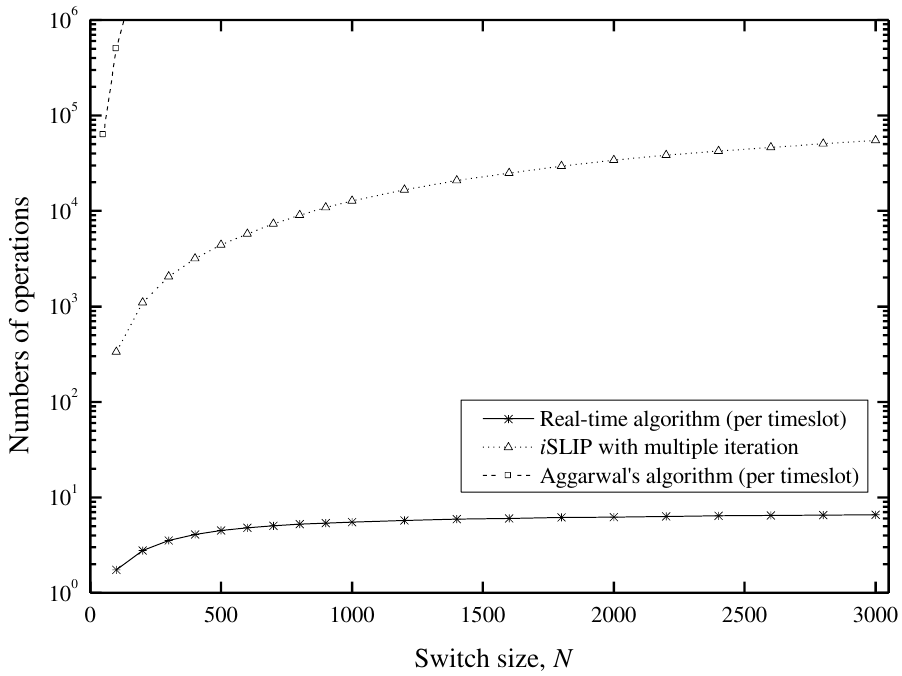}
\caption{Running time comparison of the scheduling algorithms.}
\label{performanceComplexity}
\end{figure}

The complexity of the on-line frame-based scheduling algorithm is mainly determined by the processing time 
of \emph{Parallel Complex Coloring}. We showed in previous subsection that the running time of \emph{Parallel Complex Coloring} is on the order of $O(\Delta\log⁡|V|)$. Since $\Delta{\approx}f$ if the selected frame size $f{\geq}O(\log⁡N)$ and $|V|=2N$, the running time of the frame-based scheduling algorithm is on the order of $O(\log^2⁡N)$, and the amortized complexity (matching per timeslot) is only $O(\log⁡N)$. The simulation results of the scheduling algorithms exhibited in Fig.~\ref{performanceComplexity} shows that the running time of our on-line scheduling algorithm is much less than the well-known \emph{i}SLIP algorithm and Aggarwal’s algorithm \cite{Focs03} for computing one matching per timeslot.

For easy references, Table~\ref{table_online} shows the comparison of our works and other related scheduling algorithms. The complexities listed in in Table~\ref{table_online} are all normalized to the running time of a matching per timeslot. Although the complexity of Aggarwal’s algorithm is on the order of $O(1)$ per vertex per timeslot \cite{Focs03}, but it is a sequential algorithm and cannot be parallelized because the coloring of one edge depends on the colors of its neighboring edges. Thus, for computing a matching per timeslot, the order of time complexity should be $O(N)$, which does not include the time spent at the local operations for refreshing the available color set and choosing the minimum one for an input/output pair for each color assignment. Therefore, the overall running time is on the order of $O(Nf)$, where the frame size $f$ is on the order of $O(N^2)$ \cite{Focs03}, for computing a matching per timeslot, which agrees with the simulation result displayed in Fig.~\ref{performanceComplexity}. 

\section{Conclusions}\label{conclusion}
In this paper, we study distributed and parallel scheduling algorithms based on a new algebraic edge-coloring method, called complex coloring, for packet switching. The proposed frame-based algorithm is highly efficient under both uniform and non-uniform traffic. Extensive simulations confirm performance analyses that our algorithm can achieve nearly 100\% throughput in the running time of $O(\log^2⁡N)$ per frame, which implies that the amortized complexity per timeslot is only $O(\log⁡N)$. The time complexity of parallel complex coloring can be substantially improved if we relax the constraint on the minimum number of colors. However, increasing the number of colors will lower the bandwidth utilization of the switch. In the future research, we will investigate this trade-off between time complexity and the number of colors.
\begin{table*}[!t]
\renewcommand{\arraystretch}{1.3}
\caption{Comparison of On-line Scheduling Algorithms for Input Queued Switches}
\label{table_online}
\centering
\begin{tabular}{c|c|c|c|c}
\hline \hline
\bfseries Research Work & \bfseries Complexity (matching per slot) & \bfseries Parallel & \bfseries Scheduling Granularity & \bfseries Methodology\\
\hline \hline
\emph{i}SLIP \cite{iSLIP} & $O(N{\log}N)$ & Yes & Slot by slot & Maximal size matching\\
\hline
\emph{i}LQF \cite{iLQF} & $O(N^2{\log}N)$ & Yes & Slot by slot & Maximal weighted matching\\
\hline
LAURA \cite{APSARA03} & $O(N{\log}^2N)$ & Yes & Slot by slot & Maximum weighted matching\\
\hline
G. Aggarwal etc. \cite{Focs03} & $O(Nf)$ & No & Frame by frame (frame size $f$: $O(N^2)$) & Greedy edge coloring\\
\hline
\bfseries Our work & $O({\log}N)$ & Yes & Frame by frame (frame size $f$: $O({\log}N)$) & Complex coloring\\
\hline \hline
\end{tabular}
\end{table*}

\appendices
\section{Graph Initialization}\label{appendix1}
In this appendix, we give a detailed description of the graph initialization with the current information of the packets in the $k^{th}$ time frame and the coloring information of the previous $(k-1)^{th}$ time frame. Let $C_{i,j}$ denote the set of colors that was assigned to the edges between $x_i$ and $y_j$ in the last properly colored bipartite graph $G_{k-1}$. Such historical coloring information, stored in each input line card, can be used to update the new consistent coloring of graph $G_k$. At the end of the variable elimination process, each input $i$ updates set $C_{i,j}$ according to the coloring results. When the initialization begins, each input and each output initialize the consistently colored bipartite graph in a distributed manner. The graph initialization procedure is given as follows.
\begin{algorithm}[h]
\caption{Graph Initialization}
\label{initAlg}
\begin{algorithmic}[1]
\FORALL{input $x_i$}
\FORALL{packet that is the $t^{th}$ packet destined to output $j$ in buffer}
\STATE
Add edge $e_t(x_i,y_j)$ and maintain it as $l_t(x_i,y_j)$ and $l_t(y_j,x_i)$
\IF{$C_{i,j}$ is not empty}
\STATE
color $l_t(x_i,y_j)$ and $l_t(y_j,x_i)$ with a color $c$ in $C_{i,j}$ and $C_{i,j}-\{c\}$
\ELSE
\STATE
keep $l_t(x_i,y_j)$ and $l_t(y_j,x_i)$ uncolored
\ENDIF
\ENDFOR
\STATE
send color information to associate vertices in $Y$
\ENDFOR
\FORALL{vertex $x_i$ and $y_j$}
\IF{there are uncolored links}
\STATE
color them one by one with an available color in $C$
\ENDIF
\ENDFOR
\STATE
Return a consistent coloring of the bipartite graph
\end{algorithmic}
\end{algorithm}
\begin{figure*}[!t]
\centering
\includegraphics[scale=1.08]{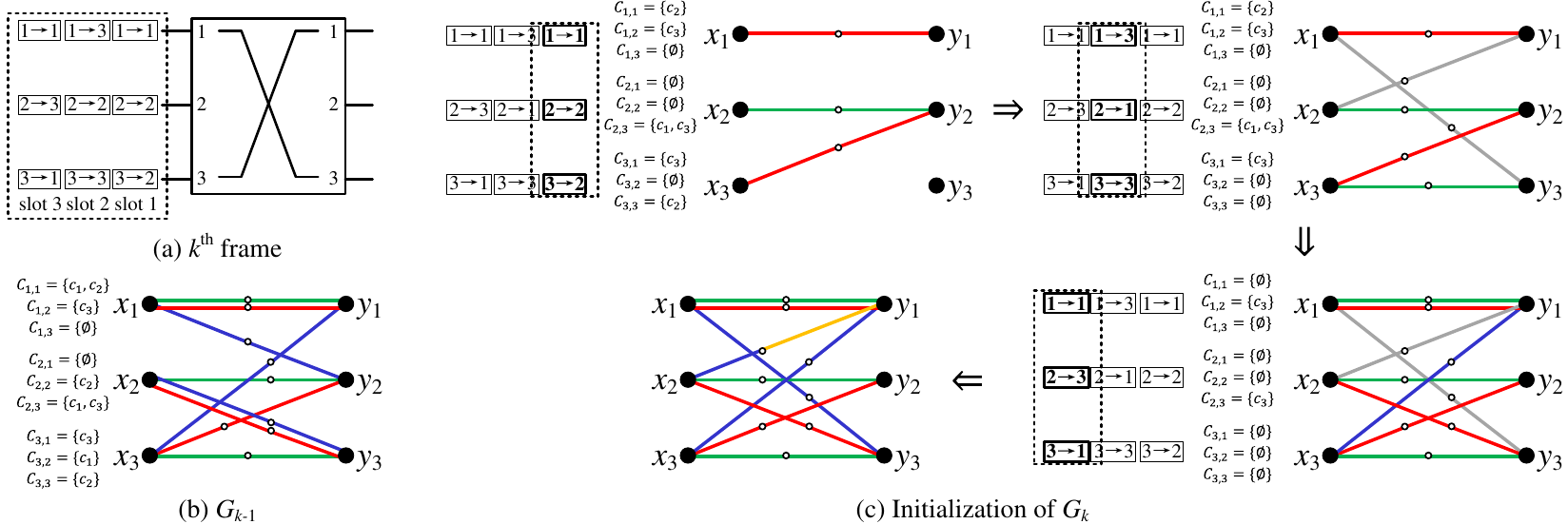}
\caption{Initialization of the consistently colored bipartite graph in a distributed manner.}
\label{FigAppendix1}
\end{figure*}

Fig.~\ref{FigAppendix1} illustrates an example of initialization of the on-line frame-based scheduling algorithm for the $k^{th}$ frame in a $3{\times}3$ switch with frame size $f=3$ and color set $C=\{c_1,c_2,c_3,c_4\}=\{r,g,b,y\}$ since $\Delta=4$. Fig.~\ref{FigAppendix1}(a) and (b) display the information of packets in $k^{th}$ time frame and the coloring information of the previous $(k-1)^{th}$ time frame, respectively. Fig.~\ref{FigAppendix1}(c) illustrates the procedure of color assignments. In this example, each input processes the packets according to their arriving orders in the frame. For instance, in the first slot of this frame, a packet destined to output 1 arrives at input 1, the corresponding edge $e_1(x_1,y_1)$ is added and assigned $c_1$, because edge $e_1(x_1,y_1)$ also existed in $G_{k-1}$ and colored by $c_1$ in the previous time frame. On the other hand, a packet destined to output 3 arrives at input 1 in the second timeslot, then edge $e_1(x_1,y_3)$ is added but remains to be colored since there is no counterpart in $G_{k-1}$. After all edges inherited from $G_{k-1}$ have been colored, then the links $l_1(x_1,y_3)$ and $l_1(y_3,x_1)$ of edge $e_1(x_1,y_3)$ will be colored respectively by input 1 and output 2 with an available color. 

\section{Selection of Frame Size $f$}\label{appendix3}
In this appendix, we discuss the selection of frame size $f$ for our scheduling algorithm to achieve a high throughput. It is easy to know that the condition $\Delta{\leq}f$ can be easily satisfied when the arrival rate is low. Therefore, we consider the boundary condition case when the traffic is fully loaded, $\lambda_i=\sum_{j}\lambda_{ij}=1$ and $\lambda_j=\sum_{i}\lambda_{ij}=1$, where $\lambda_{ij}$ is the average traffic rate from input $i$ destined to output $j$. Each input has $f$ packets to be switched in a frame, but the total number of packets that are destined for each output is a random variable even under uniform address assumption. Suppose the number of packets targeted for output $j$ is $X_j$, and let $\Delta=\max_j⁡\{X_j\}$, which means the switch needs $\Delta$ timeslots to clean up all the packets of this frame. It is easy to show that a 100\% throughput can be achieved if $\Delta{\leq}f$. If $\Delta>f$ occurs due to the randomness of the output addresses, then the transmission of some packets will be deferred. It follows that the throughput is bounded by $fN/\Delta N$ in this case. Thus, to achieve high throughput, it is necessary to maximize the factor $f/\Delta$. According to the central limit theorem \cite{ross2014}, we show in the following that $\Delta{\approx}f$ is achievable if $f{\geq}O(\log⁡N)$, provided that the number of ports of switch $N{\gg}1$. 

We assume that the addresses are uniformly distributed and the port count of switch $N{\gg}1$. Let $X_{i,j}(t)$ be the number of packets that arrive at input $i$ in the $t^{th}$ timeslot and will go to output $j$, and $X_j(t)=\sum_{i}X_{i,j}(t)$. Thus, the total number of packets that are destined for output $j$ within a frame is given by $X_j=\sum_{t=1}^{f}X_{j}(t)$. 

Under the uniform address assumption, $X_{1,j}(t)$, $X_{2,j}(t)$, $\dots$, $X_{N,j}(t)$ are i.i.d. Bernoulli random variables with parameter $1/N$, and $X_j(1)$, $X_j(2)$, $\dots$, $X_j(f)$ are i.i.d. binomial random variables, of which the expectation is 1 and the variance is $1-1/N{\to}1$ as $N{\gg}1$. It follows from the central limit theorem that
\begin{equation}
\frac{X_j-\mu}{\sigma}\to N(0,1),
\end{equation}
where $\mu=f$ and $\sigma^2=f$

Given that $X_1$, $X_2$, $\dots$, $X_N$ are i.i.d. random variables. Our goal is to achieve a high throughput $f/\Delta$. According to the result presented in \cite{leadbetter12}, the asymptotic distribution of $\Delta$ is Type I of extreme value distribution given as follows:
\begin{equation}
\Pr\{\Delta{\leq}x\}=\exp[-e^{-a_N(x-b_N)}],
\label{deltaDistribution}
\end{equation}
where
\begin{equation}
a_N=\frac{\sqrt{2\ln N}}{\sigma}, b_N=\sigma \sqrt{2\ln N}-\sigma \frac{\ln{\ln N}+\ln{4\pi}}{2\sqrt{2\ln N}}+\mu.
\label{coefficient}
\end{equation}
To achieve a targeted throughput requirement $\eta$ with a high probability $1-\varepsilon$, where $0{\ll}\eta<1,0<\varepsilon{\ll}1$, from (\ref{deltaDistribution}), we have to ensure that
\begin{equation}
\Pr\{\frac{f}{\Delta}{\geq}\eta\}=\Pr\{\Delta{\leq}\frac{f}{\eta}\}=\exp[-e^{-a_N(\frac{f}{\eta}-b_N)}]{\geq}1-\varepsilon,
\label{f/delta}
\end{equation}
Substituting (\ref{coefficient}) into (\ref{f/delta}), we obtain
\begin{align}
f&{\geq}(\frac{\eta}{1-\eta})^2{\times}\frac{1}{2\ln N} \nonumber \\
&{\times}\left[-\ln{\ln{\frac{1}{1-\varepsilon}}}+2\ln N-\frac{1}{2}(\ln{\ln N}+\ln{4\pi})\right]^2.
\label{f/delta}
\end{align}

This result clearly shows that a targeted throughput $\eta$ can be achieved with very high probability under the uniform traffic pattern if $f{\geq}O(\log⁡N)$. It also should be noted that achieving a throughput $\eta=1$ requires the frame size $f{\rightarrow}\infty$.

\begin{figure}[h]
\centering
\includegraphics[scale=0.93]{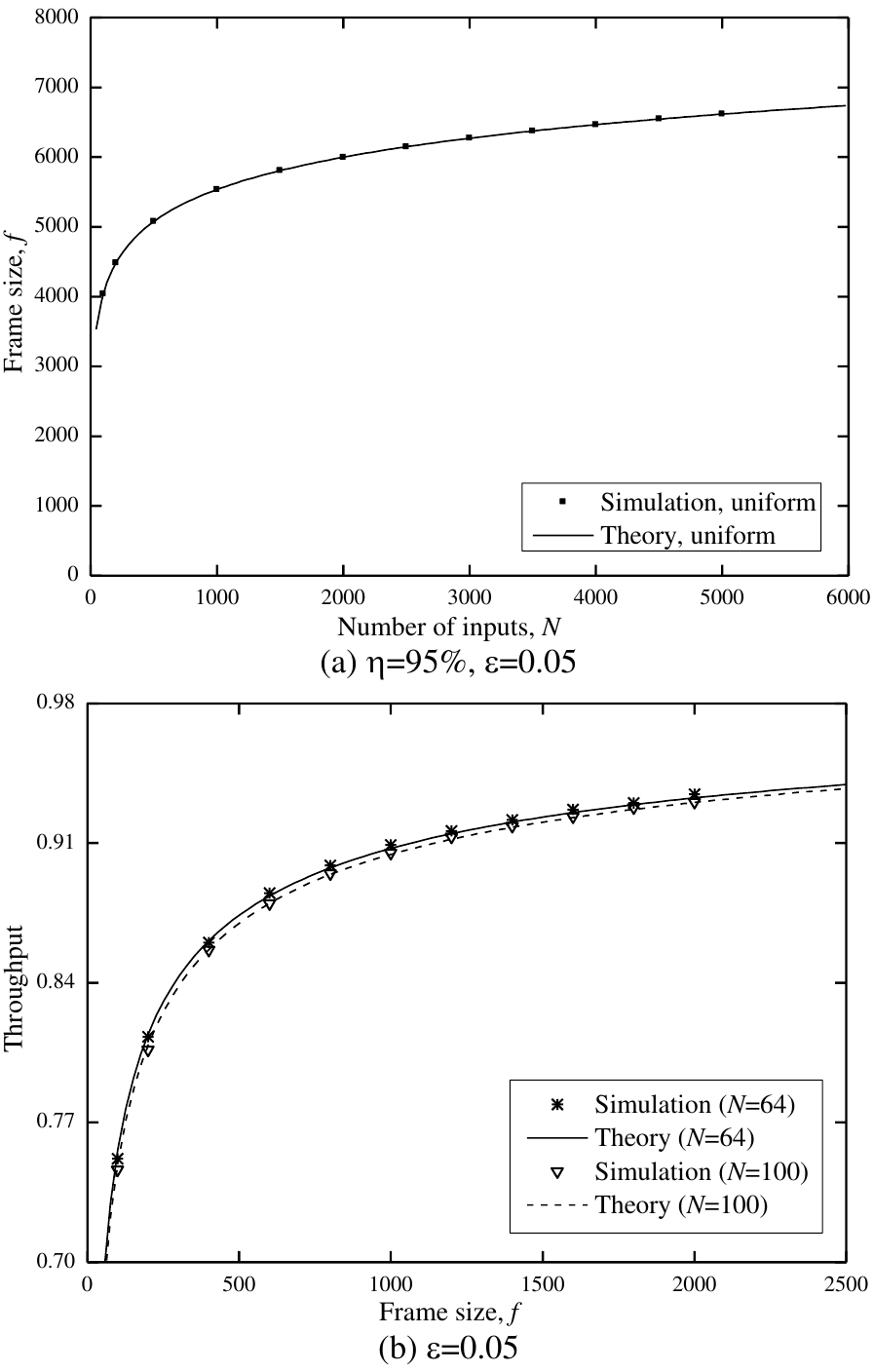}
\caption{Selection of the frame size under uniform address assumption.}
\label{figAppendix3_1}
\end{figure}

We verify the theoretical result via simulation with a fixed $\varepsilon=0.05$. For each frame size $f$, we randomly generate 10000 frames, and calculate threshold $\eta$ such that there are more than 9500 frames with $f/\Delta>\eta$. Fig.~\ref{figAppendix3_1}  shows that the theoretical results agree with the simulation results for both $N=64$ and $N=100$.

\begin{figure}[!t]
\centering
\includegraphics[scale=0.93]{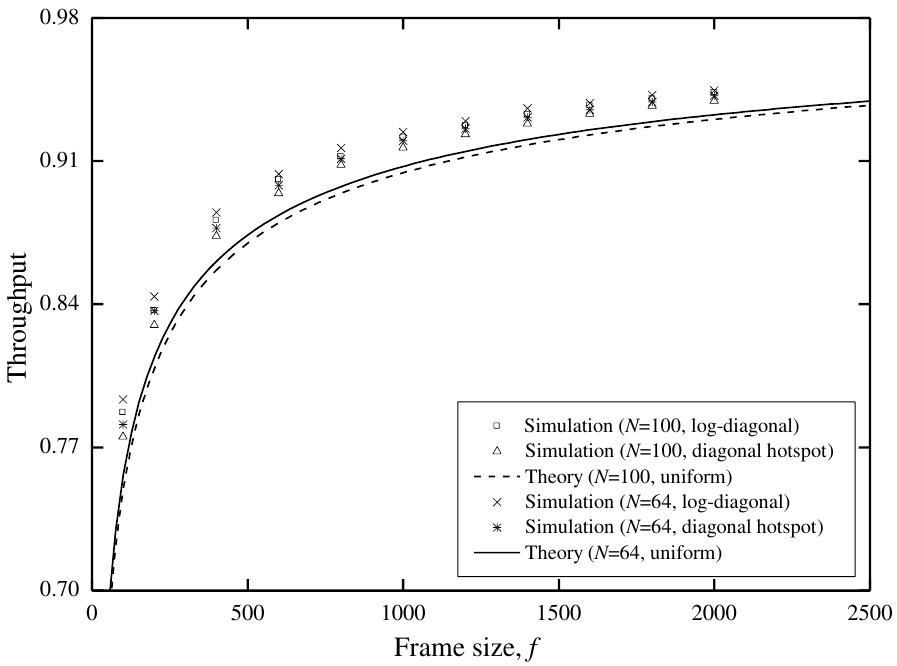}
\caption{Simulation results of the selected frame size under non-uniform address assumption with fixed $\varepsilon=0.05$.}
\label{figAppendix3_2}
\end{figure}

In comparison with the theoretical result of uniform traffic, our simulation results reveal the fact that higher throughput can be achieved under non-uniform traffic. As Fig.~\ref{figAppendix3_2} shows, higher throughputs are achieved in simulations under both types of non-uniform traffic patterns, diagonal hotspot traffic \cite{APSARA03}, and log-diagonal traffic \cite{frameMatching}. Intuitively, non-uniform traffic is more likely to concentrate on one or a few destination outputs such that the randomness of the output address of the packets is reduced. Hence, with the same frame size, throughput performance under non-uniform traffic outperforms that under uniform traffic. Therefore, the frame size selected according to requirement (\ref{f/delta}) remains valid for non-uniform input traffic.


%





\ifCLASSOPTIONcaptionsoff
  \newpage
\fi



%



\bibliographystyle{IEEEtran}
\bibliography{IEEEabrv,IEEEbib}

\end{document}